\documentclass{emulateapj}

\def\la{\mathrel{\mathchoice {\vcenter{\offinterlineskip\halign{\hfil
$\displaystyle##$\hfil\cr<\cr\sim\cr}}}
{\vcenter{\offinterlineskip\halign{\hfil$\textstyle##$\hfil\cr
<\cr\sim\cr}}}
{\vcenter{\offinterlineskip\halign{\hfil$\scriptstyle##$\hfil\cr
<\cr\sim\cr}}}
{\vcenter{\offinterlineskip\halign{\hfil$\scriptscriptstyle##$\hfil\cr
<\cr\sim\cr}}}}}

\def\ga{\mathrel{\mathchoice {\vcenter{\offinterlineskip\halign{\hfil
$\displaystyle##$\hfil\cr>\cr\sim\cr}}}
{\vcenter{\offinterlineskip\halign{\hfil$\textstyle##$\hfil\cr
>\cr\sim\cr}}}
{\vcenter{\offinterlineskip\halign{\hfil$\scriptstyle##$\hfil\cr
>\cr\sim\cr}}}
{\vcenter{\offinterlineskip\halign{\hfil$\scriptscriptstyle##$\hfil\cr
>\cr\sim\cr}}}}}

\def\kms {\hbox{${\rm km\, s}^{-1}$}} 
\def\cmsq  {$\hbox{{\rm cm}}^{-2}$}    

\def\percc {$\hbox{{\rm cm}}^{-3}$}    
\def\MOLH {\hbox{${\rm H}_2$}}  
\def\MOLN {\hbox{${\rm N}_2$}}  
\def\AMM {\hbox{${\rm NH}_{3}$}} 
\def\HCOP {\hbox{${\rm HCO}^+$}}      
\def\DCOP {\hbox{${\rm DCO}^+$}}    
\def\NTHP {\hbox{${\rm N}_{2}{\rm H}^{+}$}}   
\def\NTDP {\hbox{${\rm N}_{2}{\rm D}^{+}$}}   
\def\HTHOP {\hbox{${\rm H}_{3}{\rm O}^{+}$}}  
\def\CEIO {\hbox{${\rm C}^{18}{\rm O}$}}   
\def\CSEO {\hbox{${\rm C}^{17}{\rm O}$}}   

\shorttitle{TMC-1C: an accreting starless core}
\shortauthors{Schnee, S. et al.}

\begin{document}

\newcommand{\cso}{C$^{17}$O}
\newcommand{\ceo}{C$^{18}$O}
\newcommand{\cs}{C$^{34}$S}
\newcommand{\dcop}{DCO$^+$}
\newcommand{\nthp}{N$_2$H$^+$}
\newcommand{\ntdp}{N$_2$D$^+$}

\title{TMC-1C: an accreting starless core} 
\author{S. Schnee\altaffilmark{1,2}, P. Caselli\altaffilmark{1,3}, A. Goodman\altaffilmark{1}, H. G. Arce\altaffilmark{4}, 
	J. Ballesteros-Paredes\altaffilmark{5}, \& K. Kuchibhotla\altaffilmark{6}} 

\email{schnee@astro.caltech.edu}

       \altaffiltext{1}{Harvard-Smithsonian Center for Astrophysics, 60 Garden
       Street, Cambridge, MA 02138}
       \altaffiltext{2}{Department of Astronomy, California Institute of Technology, 
       MC 105-24 Pasadena, CA 91125}
       \altaffiltext{3}{INAF - Osservatorio Astrofisico di Arcetri, Largo E. Fermi 5, 
       50125 Firenze, Italy}
       \altaffiltext{4}{Department of Astrophysics, American Museum of Natural 
       History, New York, NY 10024}
       \altaffiltext{5}{Centro de Radioastronom\'ia y Astrof\'isica, UNAM. Apdo. 
       Postal 72-3 (Xangari), Morelia, Michoc\'an 58089, M\'exico}
       \altaffiltext{6}{Harvard Medical School, 25 Shattuck Street, Boston, MA 02115}

\begin{abstract}
We have mapped the starless core TMC-1C in a variety of molecular
lines with the IRAM 30m telescope.  High density tracers show clear
signs of self-absorption and sub-sonic infall asymmetries are present
in \NTHP (1--0) and \DCOP (2--1) lines. The inward velocity profile in
\NTHP (1--0) is extended over a region of about 7,000 AU in radius
around the dust continuum peak, which is the most extended
``infalling'' region observed in a starless core with this tracer.
The kinetic temperature ($\sim 12$~K) measured from \CSEO \ and \CEIO
\ suggests that their emission comes from a shell outside the colder
interior traced by the mm continuum dust.  The \CEIO(2--1) excitation
temperature drops from 12~K to $\simeq$10~K away from the center. This
is consistent with a volume density drop of the gas traced by the
\CEIO\ lines, from $\simeq$4$\times$10$^4$ \percc\ towards the dust
peak to $\simeq$6$\times$10$^3$ \percc\ at a projected distance from
the dust peak of 80\arcsec\ (or 11,000 AU).  The column density
implied by the gas and dust show similar \nthp\ and CO depletion
factors ($f_D \le 6$).  This can be explained with a simple scenario
in which: (i) the TMC--1C core is embedded in a relatively dense
environment ($n(\MOLH )$ $\simeq$ 10$^4$ \percc), where CO is mostly
in the gas phase and the \NTHP \ abundance had time to reach
equilibrium values; (ii) the surrounding material (rich in CO and
\NTHP ) is {\it accreting} onto the dense core nucleus; (iii) TMC-1C
is older than 3$\times 10^5$ yr, to account for the observed abundance
of \NTHP\ across the core ($\simeq$10$^{-10}$ w.r.t. \MOLH ); and (iv)
the core nucleus is either much younger ($\simeq$ 10$^4$ yr) or
``undepleted'' material from the surrounding envelope has fallen
towards it in the past 10,000 yr.
\end{abstract}


\section{Introduction}
Dense starless cores in nearby low-mass star-forming regions such as
Taurus represent the simplest areas in which to study the initial
conditions of star formation.  The dominant component of starless
cores, H$_2$, is largely invisible in the quiescent interstellar
medium, so astronomers typically rely on spectral line maps of trace
molecules and continuum observations of the thermal emission from dust
to derive their kinematics and physical state.  However, it is now
well established that different species and transitions trace
different regions of dense cores, so that a comprehensive multi--line
observations, together with detailed millimeter and sub--millimeter
continuum mapping are required to understand the structure and the
evolutionary status of an object which will eventually form a
protostar and a protoplanetary system.

Previous studies of starless cores in Taurus, as well as other nearby
star-forming regions, have shown that the relative abundance of many
molecules varies significantly between the warmer, less dense
envelopes and the colder, denser interiors (see Ceccarelli et al. 2006
and Di Francesco et al. 2006 for detailed reviews on this topic).  For
instance, \citet{Caselli99}, \citet{Bergin02} and \citet{Tafalla04}
have shown that carbon-bearing species such as \cso, \ceo, \cs, and CS
are largely absent from the cores L1544, L1498 and L1517B at densities
larger than a few 10$^4$ cm$^{-3}$, while nitrogen-bearing species
such as \nthp\ and ammonia are preferentially seen at high densities.
The chemical variations within a starless core are likely the result
of molecular freeze--out onto the surfaces of dust grains at high
densities and low temperatures, followed by gas phase chemical
processes, which are profoundly affected by the abundance drop of
important species, in particular CO \citep[see e.g.][]{Dalgarno84,
Bergin97, Taylor98, Aikawa05}.

In the past few years it has also been found that not all starless
cores show a similar pattern of molecular abundances and physical
structure.  Indeed, there is a subsample of starless cores (often
called pre-stellar cores), which are particularly centrally
concentrated, that shows kinematic and chemical features typical of
evolved objects on the verge of star formation.  These features
include large values of CO depletion and deuterium fractionation, and
evidence of ``central'' infall, i.e.  presence of infall asymmetry in
high density tracers in a restricted region surrounding the mm
continuum dust peak \citep{Williams99, Caselli02b, Redman02, Crapsi05,
Williams06}.  It is interesting that not all {\it physically} evolved
cores show {\it chemically} evolved compositions, as shown by
\citet{Lee03} and \citet{Tafalla04b}. It is thus important to study in
detail a larger number of cores to understand what is causing the
chemical differentiation in objects with apparently similar physical
ages. This is why we decided to focus our attention on TMC--1C, a
dense core in Taurus, with physical properties quite similar to the
prototypical pre--stellar core L1544 (also in Taurus), to study
possible differences and try to understand their nature.

TMC--1C is a starless core in the Taurus molecular cloud, with a
distance estimated at 140 pc \citep{Kenyon94}.  In a previous study,
we have shown that TMC--1C has a mass of 6 M$_\odot$ within a radius
of 0.06 pc from the column density peak, which is a factor of two
larger than the virial mass derived from the \nthp (1--0) line width,
and we have shown that there is evidence for sub-sonic inward motions
\citep{Schnee05a} as well as a velocity gradient consistent with solid
body rotation at a rate of 0.3 km s$^{-1}$ pc$^{-1}$
\citep{Goodman93}.  TMC--1C is a coherent core with a roughly constant
velocity dispersion, slightly higher than the sound speed, over a
radius of 0.1 pc \citep{Barranco98, Goodman98}.  Using SCUBA and MAMBO
bolometer maps of TMC-1C at 450, 850 and 1200 \micron, we have mapped
the dust temperature and column density and shown that the dust
temperature at the center of the core is very low ($\sim$6 K)
\citep{Schnee06b}.
	
In order to disentangle the physical and chemical information that can
be gleaned from a combination of gas and dust observations of a dense
core, we have now mapped TMC-1C at three continuum wavelengths
\citep{Schnee06b} and seven molecular lines.  In
Sec.~\ref{OBSERVATIONS}, continuum and line observations are
described. Spectra and maps are presented in Sec.~\ref{RESULTS}. The
analysis of the data, along with the discussion, has been divided in
three parts: kinematics, including line width variations across the
cloud, velocity gradients and inward velocities, is in
Sec.~\ref{ANALYSIS_I}; gas and dust column density and temperature are
in Sect.~\ref{ANALYSIS_II}; molecular depletion and chemical processes
are discussed in Sect.~\ref{ANALYSIS_III}. A summary can be found in
Sect.~\ref{SUMMARY}.

\section{Observations} \label{OBSERVATIONS}
\subsection{Continuum} \label{CONTINUUM}

To map the density and temperature structure of TMC--1C, we have
observed thermal dust emission at 450 and 850 \micron\ with SCUBA and
at 1200 \micron\ with MAMBO--2.

\subsubsection{SCUBA}
We observed a 10\arcmin$\times$10\arcmin\ region around TMC-1C using
SCUBA \citep{Holland99} on the JCMT at 450 and 850 \micron.  We used
the standard scan-mapping mode, recording 450 and 850 \micron\ data
simultaneously \citep{Pierce-Price00,Bianchi00}.  Chop throws of
30\arcsec, 44\arcsec\ and 68\arcsec\ were used in both the right
ascension and declination directions.  The resolution at 450 and 850
\micron\ is 7.5\arcsec\ and 14\arcsec\ respectively.  The absolute
flux calibration is $\sim$12\% at 450 \micron\ and $\sim$4\% at 850
\micron.  The noise in the 450 and 850 \micron\ maps are 13 and 9
mJy/beam, respectively.  The data reduction is described in detail in
\citet{Schnee06b}.

\subsubsection{MAMBO-2}
Kauffmann et al. (in prep.) used the MAMBO-2 array \citep{Kreysa99} on
the IRAM 30--meter telescope on Pico Veleta (Spain) to map TMC--1C at
1200 \micron.  The MAMBO beam size is 10\farcs7.  The source was
mapped on-the-fly, chopping in azimuth by 60\arcsec\ to 70\arcsec\ at
a rate of 2 Hz.  The absolute flux calibration is uncertain to
$\sim$10\%, and the noise in the 1200 \micron\ map is 3 mJy/beam.  The
data reduction is described in detail in Kauffmann et al. (in prep.).

\subsection{Spectral Line} \label{SPECTRA}

We have used the IRAM 30-m telescope to map out emission from several
spectral lines in order to understand the kinematic and spectral
structure of TMC--1C.  In November 1998, we mapped the spectral line
maps of the \cso (1--0), \cso (2--1), \ceo (2--1), \cs (2--1), \dcop
(2--1), \dcop (3--2), \nthp (1--0) transitions.  The inner 2\arcmin\
of TMC-1C were observed with 20\arcsec\ spacing in frequency-switching
mode, and outside of this radius the data were collected with
40\arcsec\ sampling.  The data were reduced using the CLASS package,
with second-order polynomial baselines subtracted from the spectra.
The system temperatures, velocity resolution, beam size and beam
efficiencies are listed in Table \ref{IRAMOBS}.

\section{Results} \label{RESULTS}
\subsection{Spectra} \label{RESULTS_SPECTRA}

\begin{figure}
\epsscale{0.8}
\plotone{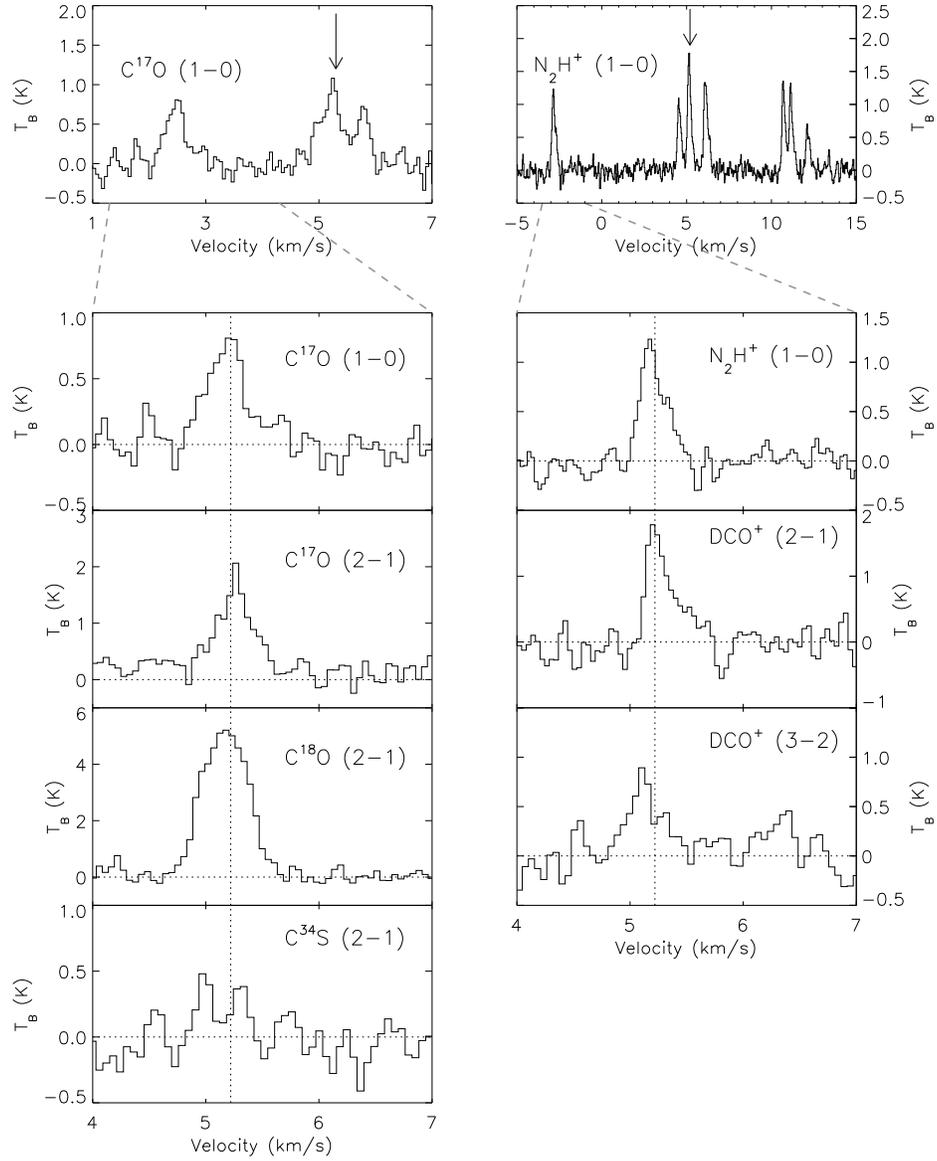}
\caption{The spectra taken at the (0,20) position, which is the peak
  of the dust emission map.  The velocity scales on the spectra
  showing hyperfine structure are only correct for the marked
  component, whereas the enlarged portion of the spectra has been
  recentered at the frequency of the isolated component.  The vertical
  dotted line shows the LSR velocity of the weakest component of the 
  \nthp (1-0) spectrum, as determined by a Gaussian fit, for comparison.
	 \label{IRAMSPEC}}
\end{figure}

The spectra taken at the peak of the dust column density map are shown
in Figure \ref{IRAMSPEC}.  The integrated intensity, velocity, line
width and RMS noise for each transition is given in Table
\ref{FITTABLE}. From the figure it is evident that self--absorption is
present everywhere, except in \CSEO \ and \CEIO \ lines.  Clear
signatures of inward motions (brighter blue peak; e.g. Myers et
al. 1997, see also Sect.~\ref{INFALL}) are only present in the high
density tracers \NTHP and \DCOP , which typically probe the inner
portion of dense cores (e.g.  Caselli et al. 2002b; Lee et
al. 2004). The C$^{34}$S(2--1) line appears to be self--absorbed at
the cloud velocity, but the spectrum is too noisy to confirm this.

To highlight the extent of the ``infall'' asymmetry in \NTHP (1--0),
Fig.~\ref{spectra_map} shows the profile of the main hyperfine
component of the \NTHP (1--0) transition (F$_1$,F = 2,3 $\rightarrow$
1,2) across the whole mapped area (see Sect.~\ref{MAPS}). We will
discuss these spectra in more detail in Sect.~\ref{INFALL}, but for
now it is interesting to see how the profile shows complex structure,
consistent with inward motion (red boxes) as well as outflow (blue
boxes) and absorption from a static envelope.

\subsection{Maps} \label{MAPS}

From Fig.~\ref{IRAMSPEC} and \ref{spectra_map}, it is clear that the
main \NTHP (1--0) hyperfine components are self-absorbed around the
dust peak position.  Therefore, a map of the \NTHP (1--0) intensity
integrated under the seven components will not reflect the \NTHP \
column density distribution.  However, the weakest component (F$_1$ F
= 10 $\rightarrow$ 11) is not affected by self-absorption, as shown in
Fig.~\ref{infall_dust_peak}, where the weakest and the main (F$_1$ F =
23 $\rightarrow$ 12) components toward the dust peak position (the
most affected by self-absorption) are plotted together for
comparison. The two hyperfines in Fig.~\ref{infall_dust_peak} have
very different profiles: the main component is blue-shifted,
suggestive of inward motions (see Sect.~\ref{INFALL}), whereas the
weak component is symmetric and its velocity centroid is red shifted
compared to the main component, indicating optically thin
emission. Given that the self-absorption is more pronounced at the
position of the dust peak, where the \NTHP (1--0) optical depth is
largest, we conclude that the weak component is likely to be optically
thin across the whole TMC--1C core.

Thus, in the case of \NTHP (1--0) self--absorption, we used the weak
hyperfine component, divided by 1/27 (its relative intensity compared
to the sum, normalized to unity, of the seven hyperfines), to
determine the \NTHP (1--0) integrated intensity, line width, and, as
shown in Sec.~\ref{ANALYSIS_II}, the \NTHP \ column density.  In this
analysis, only spectra with signal to noise (S/N) for the weak
component $>$ 3 have been considered.  In all other cases, \NTHP
(1--0) did not show signs of self--absorption and a normal integration
below the 7 hyperfines has been performed.

Based on hyperfine fits to the \cso (1--0) transition and a comparison
of the relative strengths of the three components, we see that the
\cso (1--0) emission is optically thin throughout TMC--1C.  Although the
noise is generally too high in the \cso (2--1) data to make
indisputable hyperfine fits, the results of such an attempt suggest
that the \cso (2--1) lines are also optically thin, which is expected
for thin \cso (1--0) emission and temperatures of $\sim$10 K. Thus,
the integrated intensity maps of \CSEO \ lines will reflect the \CSEO
\ column density distribution.

In order to estimate the optical depth of the \ceo (2--1) lines, we
compare the integrated intensity of \ceo (2--1) to that of \cso
(2--1).  If both lines are thin, then the observed ratio should be
equal to the cosmic abundance ratio, $R_{18,17} \equiv$
[$^{18}$O]/[$^{17}$O] = 3.65 \citep{Wilson94, Penzias81}.  We observe
that the ratio of the integrated intensities $R_{18,17} = 2.3 \pm
0.9$, which corresponds to an optical depth of the \ceo (2--1) line of
$\tau_{18} \simeq 1.5$.

Integrated intensity maps of \cso (1--0), \cso (2--1), \ceo (2--1),
\dcop (2--1) and \nthp (1--0) are shown in Fig.~\ref{INTMAPS}.  Note
that the \nthp\ integrated intensity map peaks right around the
position of the dust column peaks, which is not true for \cso\ and
\ceo.  We do not present integrated intensity maps of \cs (2--1) or
\dcop(3--2), which have lower signal to noise.

\section{Analysis. I. Kinematics} \label{ANALYSIS_I}

\subsection{Line Widths}

Ammonia observations have shown that TMC--1C is a coherent core,
having a constant line width across the core at a value slightly
higher than the thermal width, and increasing outside the ``coherent''
radius, $\sim$0.1 pc \citep{Barranco98}.  Our \nthp\ observations of
TMC--1C show that the line width remains constant, at a value $\sim$2
times higher than the thermal line width, over the entire core (see
Fig.~\ref{FWHMBINS} for a map of the line width and a plot of line
width vs. radius), though the dispersion in the \nthp\ line width is
very large.  This result is in agreement with the NH$_3$ observations
of TMC--1C, and is not consistent with the decreasing \nthp\ and
\ntdp\ linewidths at larger radii seen in L1544 and L694-2, which in
other important ways (density and temperature structure, velocity
asymmetry seen in \NTHP (1--0)) closely resemble TMC--1C.  To make
sure that the lack of correlation of the \NTHP (1--0) line width is
not due to geometric effects, considering the elongated structure of
TMC--1C, we also plotted the \NTHP (1--0) line width as a function of
antenna temperature, and found similar results.

This behavior may be due to the coherence of the central portion of
the core, which has nearly constant length along the line of sight,
and thus the velocity dispersion comes from regions of the core that
have similar scales (see Fig. 4 in Goodman et al. 1998).  Cores formed
by compressions in a supersonic turbulent flow naturally develop these
regions of constant length at their centers \citep{Klessen05}.
Another reason that the \nthp\ line widths appear constant across the
cloud could be the different ``infall'' velocity profile, with the
velocity peaking farther away from the dust peak than in L1544 and
L694-2 (though the projected velocity would still be at its maximum at
the dust peak).  In the case of L1544, \citet{Caselli02b} showed that
the \NTHP (1--0) line profile is consistent with the \citet{Ciolek00}
model at a certain time in the cloud evolution, where the ``infall''
velocity profile peaks at a radius of about 3,000 AU (see also Myers
2005 for alternative models with similar radial velocities).  Indeed,
in Sec.~\ref{INFALL} we show that the extent of the asymmetry seen in
\NTHP (1--0) suggests that the peak of the inward motions is at about
7,000 AU, so that one does not expect to see broader line widths
within this radius.  In fact, the binned data in Fig.~\ref{FWHMBINS}
show a hint of a peak at about 50\arcsec\ (7000 AU at the distance of
Taurus).

In order to compare the thermal and non-thermal line widths in TMC-1C,
we assume that the gas temperature is equal to 10 K and use the
formulae:

\begin{equation}
\Delta v_T = \sqrt{\left(\frac{8 \ln(2) \, k T_g}{\mu m_H}\right)}
\end{equation}
\begin{equation}
\Delta v_{NT} = \sqrt{(\Delta v_{obs})^2 - (\Delta v_T)^2}
\end{equation}
where $\mu$ is the molecular weight of the species and $m_H$ is the
mass of hydrogen.  

Figure \ref{FWHMRATIO} shows the non-thermal line width plotted
against the thermal line width at the position of the dust column
density peak.  To allow a fair comparison, all the data in the figure
have been first spatially smoothed at the same resolution of the \AMM\
map (1\arcmin ). Although \cso, \ceo, \dcop\ and \nthp\ all have
similar molecular weights, they have significantly different values
for their non-thermal line widths.  The thermal line width is much
smaller than the non-thermal line width for the molecules \cso\ and
\ceo, while the ratio is closer to unity for \dcop\ and \nthp.  This
suggests that the isotopologues of CO are tracing material at larger
distances from the center, with a larger turbulent line width, than
are \dcop\ and \nthp, which presumably are tracing the higher density
material closer to the center of TMC-1C.  The \ceo (2--1) line is
slightly thick, and this is probably the reason of its slightly larger
line width when compared to the thin \cso (2--1) line, as shown in
Figure \ref{FWHMRATIO}.  For each transition observed, we see no clear
correlation between the observed line width and the thermal line width
(and therefore with temperature, column density and distance from the
peak column density, see Section \ref{DUSTCOLUMN}).  As in the study
of depletion, the lower signal to noise in \cs (2--1), \dcop (2--1) and
\dcop (3--2) make any possible trends between $\Delta v$ and $\Delta
v_{thermal}$ more difficult to determine.  From Fig.~\ref{FWHMRATIO}
we note that NH$_3$ and \nthp \ have similar non-thermal line widths,
which makes sense given that \NTHP \ and \AMM \ are expected to trace
similar material (e.g. Benson et al. 1998). However, this result is in
contrast with the findings of \citet{Tafalla04} who found narrower
\AMM\ line widths towards L1498 and L1517B. 

We finally note that the line widths that we measure in \cso\ and
\ceo\ are larger than the \nthp\ linewidths throughout TMC--1C, 
which contrasts with the results seen in \ceo\ and \nthp\ in B68
\citep{Lada03}.  This is consistent with the fact that TMC--1C, unlike
B68, is embedded in a molecular cloud complex and it is not an
isolated core.  Thus, CO lines in TMC--1C also trace the (lower
density and more extended) molecular material, part of the Taurus
complex, where larger ranges of velocities are present along the line
of sight.

\subsection{Velocity Gradients}
\label{svel}

In order to study the velocity field of TMC-1C, we determine the
centroid velocities for \cso (2--1), \ceo (2--1), \cs (2--1), 
\dcop (2--1) and \dcop (3--2) with Gaussian fits.  The centroid 
velocities of the \cso (1---0) and \nthp (1--0) lines are determined
by hyperfine spectral fits.  For those \nthp\ spectra that show
evidence of self-absorption, the velocity is derived from a Gaussian
fit to the thinnest component.  The velocity gradient at each position
is calculated by fitting the velocity field with the function:

\begin{equation} \label{VLSREQ}
v_{lsr} = v_o + \frac{dv}{ds} \Delta \alpha \cos{\theta} +
                \frac{dv}{ds} \Delta \delta \sin{\theta}
\end{equation}
where $v_o$ is the bulk motion along the line of sight, $\Delta
\alpha$ and $\Delta \delta$ are RA and DEC offsets from the position
of the central pixel, $dv/ds$ is the magnitude of the velocity
gradient in the plane of the sky, and $\theta$ is direction of the
velocity gradient.  The fit to the velocity gradient is based on
fitting a plane through the position--position velocity cube as in
\citet{Goodman93} (for the ``total'' gradient across the cloud) and in
\citet{Caselli02c} (for the ``local'' gradient at each position).  The
fit for the ``total'' velocity gradient gives a single direction and
magnitude for the entire velocity field analyzed.  The ``local''
velocity gradient is calculated at each position in the spectral line
maps based on the centroid velocities of the center position and its
nearest neighbors, with the weight given to the neighbors decreasing
exponentially with their distance from the central position.

Analysis of ammonia observations with $\sim60$\arcsec\ resolution of
TMC--1C indicate an overall velocity gradient of 0.3 km s$^{-1}$
pc$^{-1}$ directed 129 degrees East of North \citep{Goodman93}.  The
velocity field that we measure in TMC--1C has spatial resolution three
times greater ($\sim$20\arcsec) than the ammonia study, and reveals a
pattern more complicated than that of solid body or differential
rotation.  The velocity fields measured by \cso (1--0), \cso (2--1) and
\ceo (2--1) are shown in Fig.~\ref{COARROWS}.  Although there is a
region that closely resembles the velocity field expected from
rotation (gradient arrows of approximately equal length pointing in
the same direction), the measured velocities vary from blue to red to
blue along a NW - SE axis.  The \nthp (1--0) velocity fields (shown in
Fig.~\ref{COARROWS}) also follow the same blue to red to blue pattern
along the NW - SE axis, but the observations cover a somewhat
different area than the CO observations, which complicates making a
direct comparison.  Taken as a whole, it is clear that there is an
ordered velocity field in portions of the TMC-1C core, and that the
lower density CO tracers ``see'' a velocity field similar to that
probed by \nthp\ lines, which trace higher density material.  In any
case, the velocity field that looks like rotation reported in
\citet{Goodman93} turns out to be more complicated when seen over a
larger area with finer resolution.  The direction and magnitude of the
velocity gradient in the region that resembles solid body rotation is
shown in Fig.~\ref{CENTERGRAD} for each transition.

\subsection{Inward Motions} \label{INFALL}

To quantify the velocity of the inward motion from the \NTHP (1--0)
line across the TMC--1C cloud, we use a simple two--layer model,
similar to that described by \citet{Myers96}.  This model assumes that
the cloud can be divided in two parts with uniform excitation
temperature ($T_{\rm ex}$, gradients in $T_{\rm ex}$ between the two
layers as in De Vries \& Myers 2005, are not considered here), line
width ($\Delta v$), optical depth ($\tau$) and LSR velocity ($V_{\rm
LSR}$) and that the foreground layer has a lower excitation
temperature. For simplicity, we also assume that the seven hyperfines
have the same $T_{\rm ex}$, which is a very rough assumption in
regions of large optical depth, as recently found by \citet{Daniel06}.
Despite of the simplicity of the model, we find good fits to the seven
hyperfine lines and determine the value of the velocity difference
between the two layers, which can be related to the ``infall''
velocity.

\begin{figure}
\epsscale{1.1}
\plotone{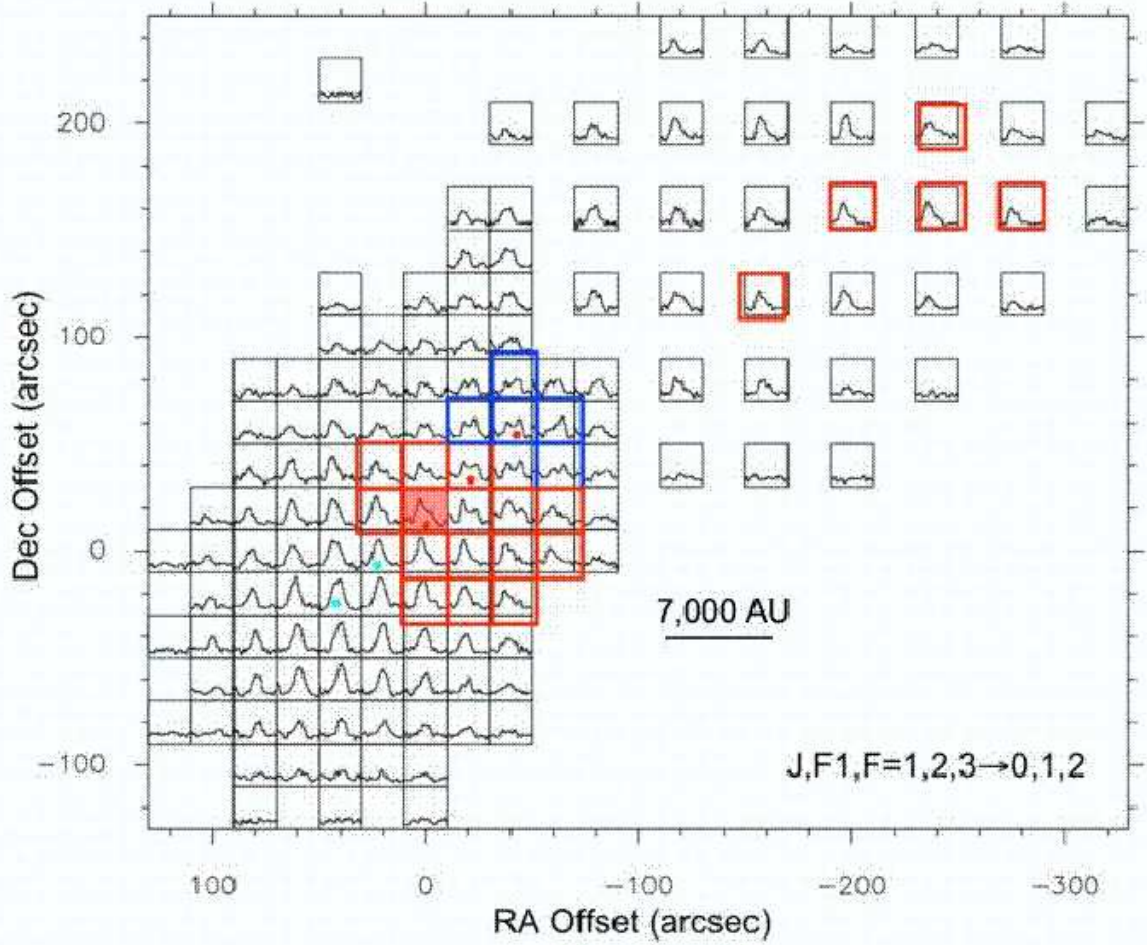}
\caption{Spectra of the F$_1$,F = 2,3 $\rightarrow$ 1,2 hyperfine
  component of \NTHP \ across the whole TMC--1C mapped region. Red
  boxes mark spectra consistent with inward motions, whereas blue
  boxes mark spectra with outflow signatures (see Sect.~\ref{INFALL}).
  Colored dots indicate those positions displayed in
  Fig.~\ref{two_layer}.  The red-filled box mark the position of the
  continuum dust emission peak. The asymmetry in the high density
  tracer \NTHP \ is observed in an extended region, larger than that
  previously found in other evolved pre--stellar cores.  See
  Fig. \ref{INTMAPS} for contours of column density based on dust
  emission.  \label{spectra_map}}
\end{figure}

\begin{figure}
\epsscale{1.0}
\plotone{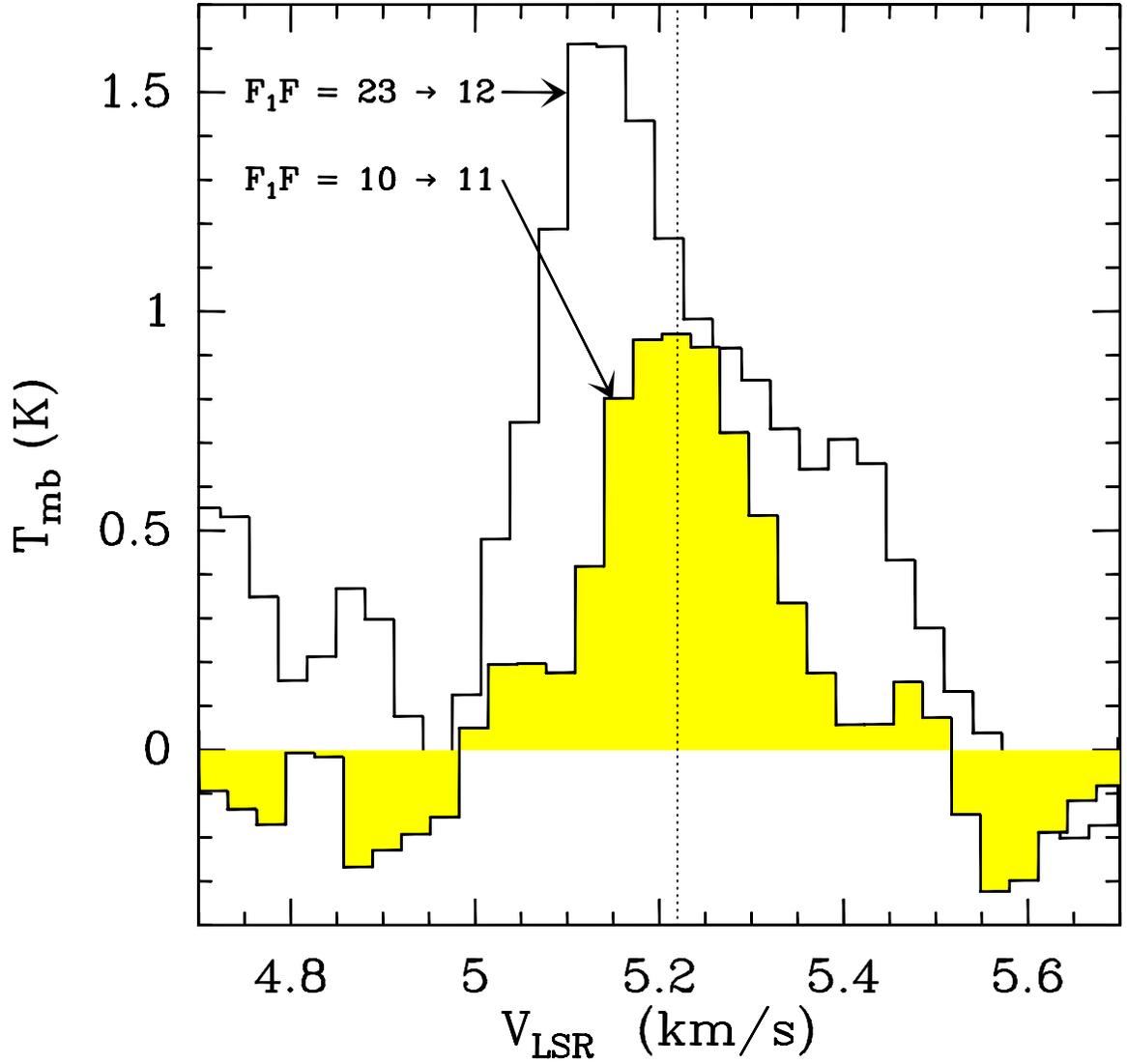}
\caption{Profile of the main F$_1$,F = 23 $\rightarrow$ 12 hyperfine
  component of \NTHP (1--0) (white histogram) superposed on the
  spectrum of the weak F$_1$F = 10 $\rightarrow$ 11 component (yellow
  histogram) at the dust peak position. Note that the weak component
  is symmetric and peaks where the absorption of the main component is
  steepest, suggesting that this component is optically thin and can
  be used to estimate the column density and line width where the main
  group of hyperfines is affected by self-absorption.
         \label{infall_dust_peak}}
\end{figure}

\begin{figure}
\epsscale{1.0}
\plotone{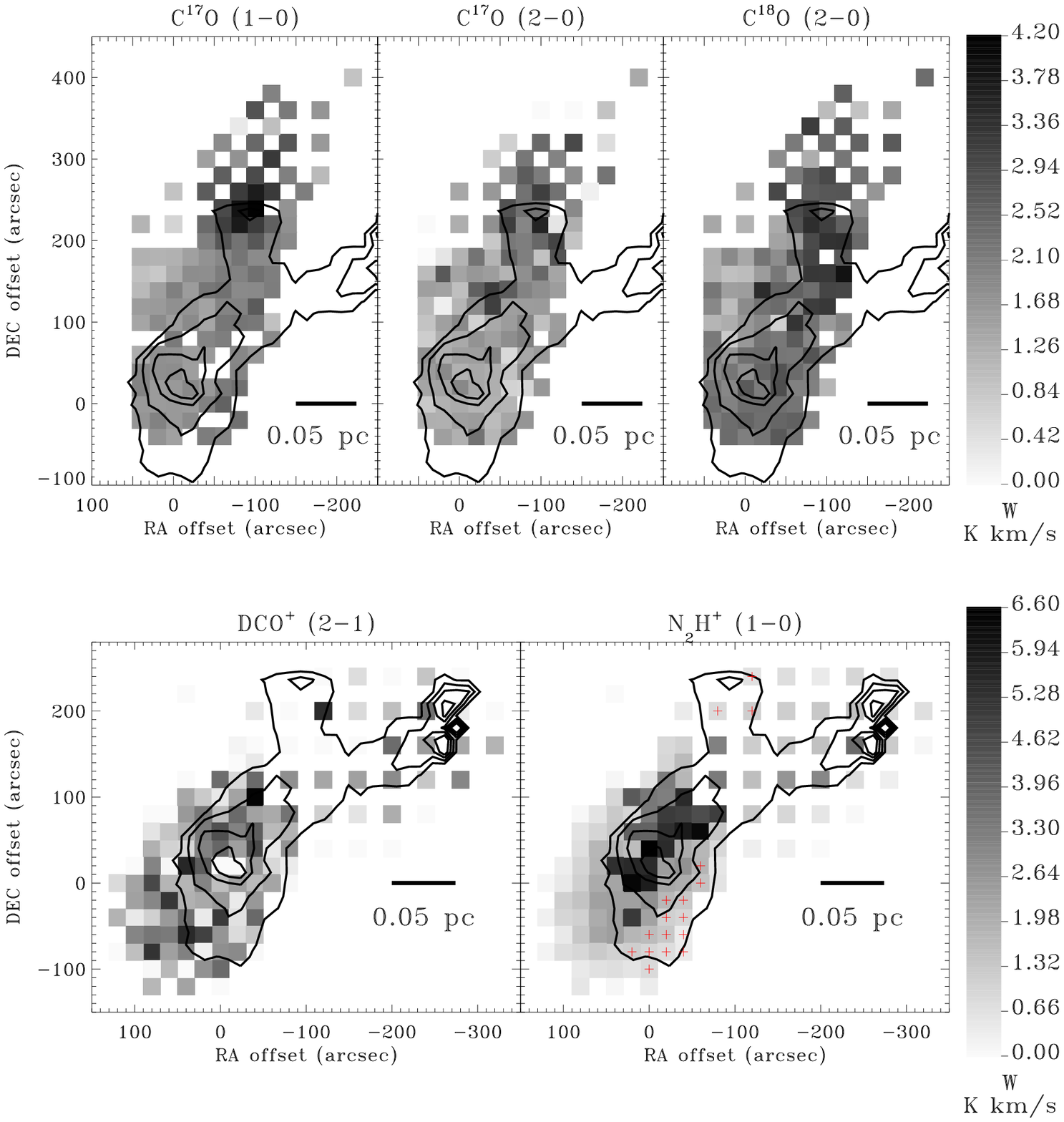}
\caption{Integrated intensity maps of ({\it top left}) \cso (1--0),
  ({\it top middle}) \cso (2--1), ({\it top right}) \ceo (2--1), ({\it
  bottom left}) \dcop (2--1) and ({\it bottom right}) \nthp
  (1--0). The \cso (1--0), \cso (2--1) and \dcop (2--1) integrated
  intensities have been scaled up by factors of 1.5, 2.0 and 5.0,
  respectively.  For all spectra showing self-absorption, the \nthp\
  integrated intensity has been calculated from the thinnest hyperfine
  component.  The contours show the visual extinction derived from the
  dust emission, at levels of $A_V$ = 10, 25, 40, and 55 mag.  The
  1$\sigma$ uncertainties in the integrated intensity are $\sim$0.05,
  $\sim$0.06, $\sim$0.10, $\sim$0.07 and $\sim$0.08 K km s$^{-1}$ in
  \cso (1--0), \cso (2--1), \ceo (2--1), \dcop (2--1) and \nthp (1--0)
  respectively.  The positions with large \nthp\ depletion and low
  column density have been marked with red crosses.  \label{INTMAPS}}
\end{figure}

\begin{figure}
\epsscale{0.8}
\plotone{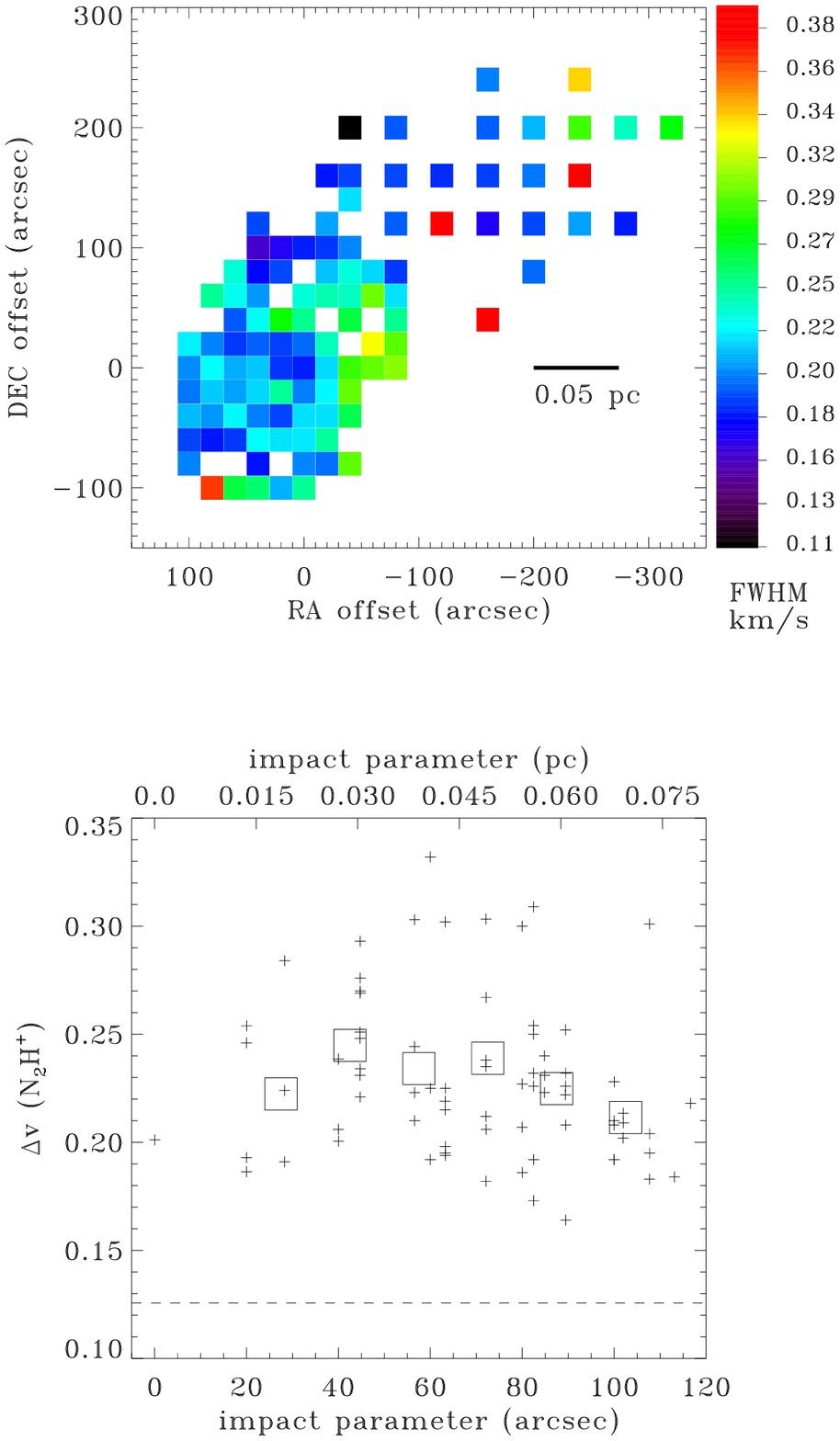}
\vspace{10 mm}
\caption{({\it Top Panel}) \nthp (1-0) observed line width across the 
TMC--1C core. ({\it Bottom Panel}) \nthp (1-0) observed line width (in
km s$^{-1}$) vs. distance from the peak of the dust column density
map.  For those positions where the self--absorption is present, we
used the width of the weak component as representative of the
intrinsic line width of the (1-0) line. The crosses show the values at
each position in the map with S/N $>$ 3, and the large squares are
averages of the data with 15\arcsec\ bins.  The horizontal dashed line
shows the thermal FWHM at 10 K for reference.  \label{FWHMBINS}}
\end{figure}

\begin{figure}
\epsscale{0.8}
\plotone{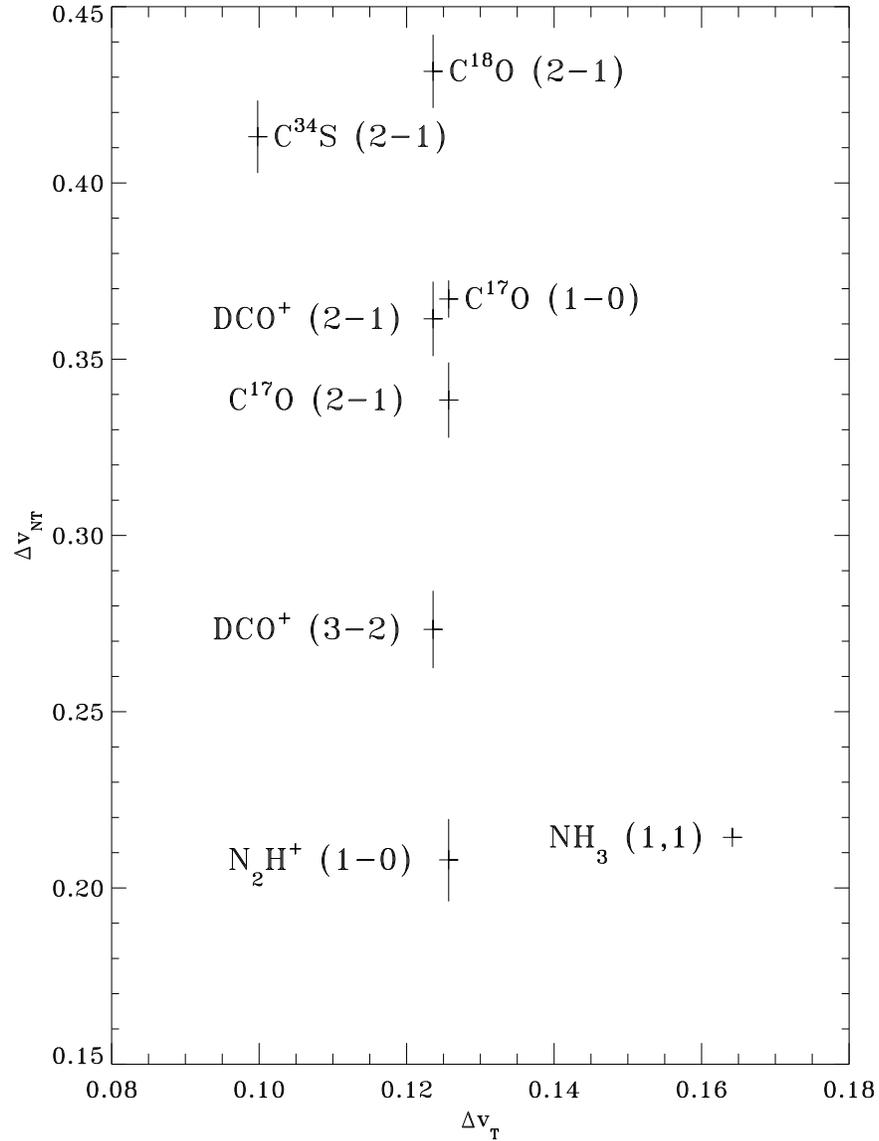}
\caption{The non-thermal line width plotted against the thermal line
  width for each transition observed, at the position of the dust
  emission peak.  The thermal line width is calculated from an assumed
  gas temperature of 10 K. The error in the non-thermal line width is
  derived using the error in the FWHM of the Gaussian fit to the line.
  The line withs of C$^{34}$S(2--1) and \DCOP (3--2) are highly 
uncertain, because of the strong absorption observed (see 
Fig.~\ref{spectra_map}) and the plotted errors do not take the 
absorption into account. 
  The NH$_3$ intrinsic line width is taken from \citet{Barranco98}.
  To make a fair comparison, all of the spectral line maps have been
  spatially smoothed to the $\sim$1\arcmin\ resolution of the NH$_3$
  map.  \label{FWHMRATIO}}
\end{figure}

\begin{figure}
\epsscale{1.0}
\plotone{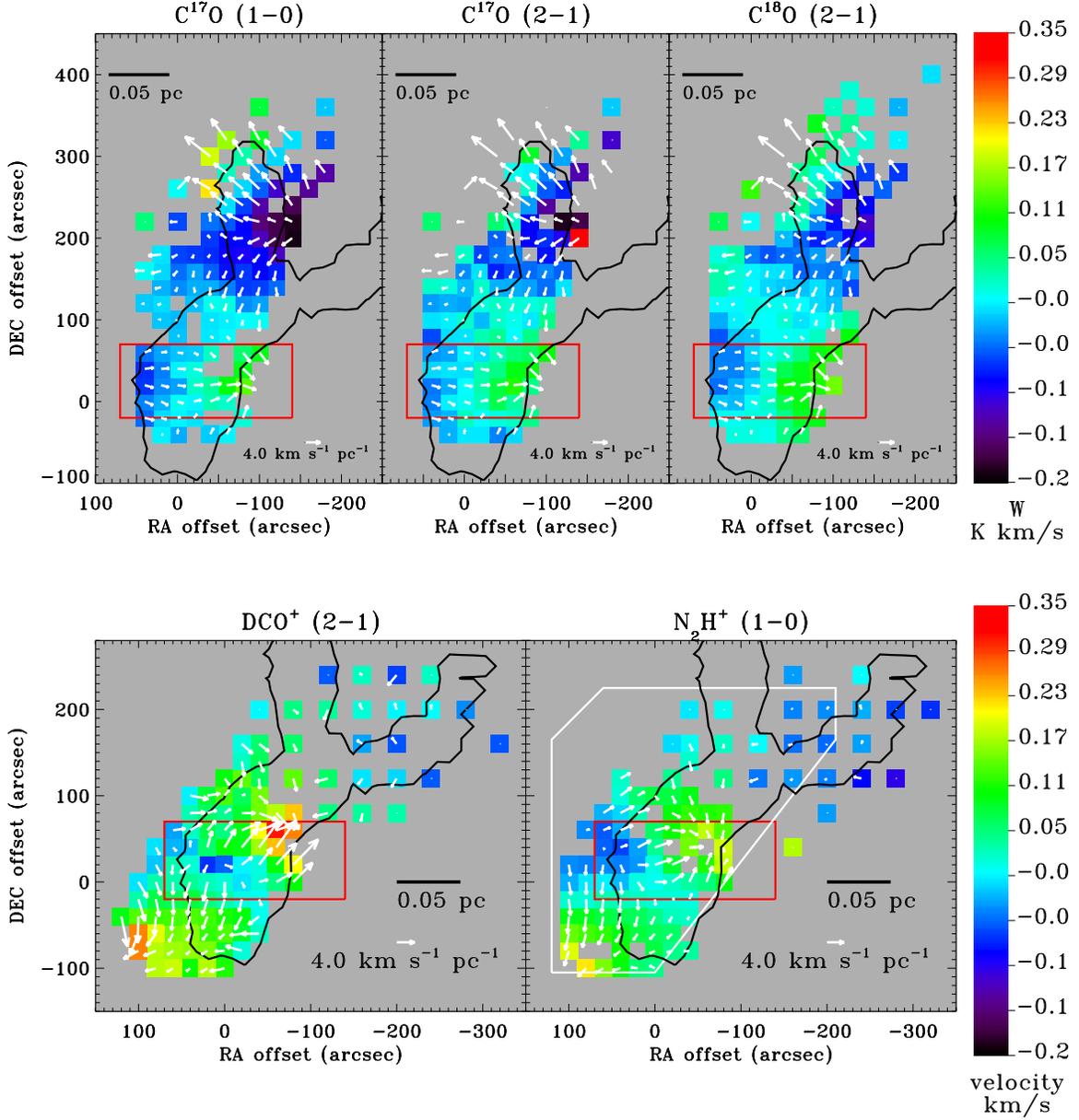}
\caption{{\it Top Panels}) 
The velocity field for \cso (1-0), \cso (2-1) and \ceo (2-1).
  The black contour shows the A$_{\rm V} = 10$~mag boundary, and the
  arrows show the direction and magnitude of the line-of-sight
  velocity gradient, pointing from blue to red.  The red box encloses
  a region with a velocity gradient similar to that of rotation as
  seen in the isotopologues of CO.  In each map the velocity shown is
  the difference between the measured VLSR at each position and the
  velocity at the (0,0) position.  The lengths of the arrows are
  proportional to the magnitude of the gradient, and is scaled to the
  labeled arrow.  ({\it Bottom Panels}) velocity field for
   \DCOP (2--1) and \nthp (1--0) in TMC-1C.  The red box encloses a
  region with a velocity gradient in isotopologues of CO that looks
  similar to that of rotation.  The white box shows the region
  observed in NH$_3$ by \citet{Barranco98}.
  \label{COARROWS}}
\end{figure}

\begin{figure}
\epsscale{1.0}
\plotone{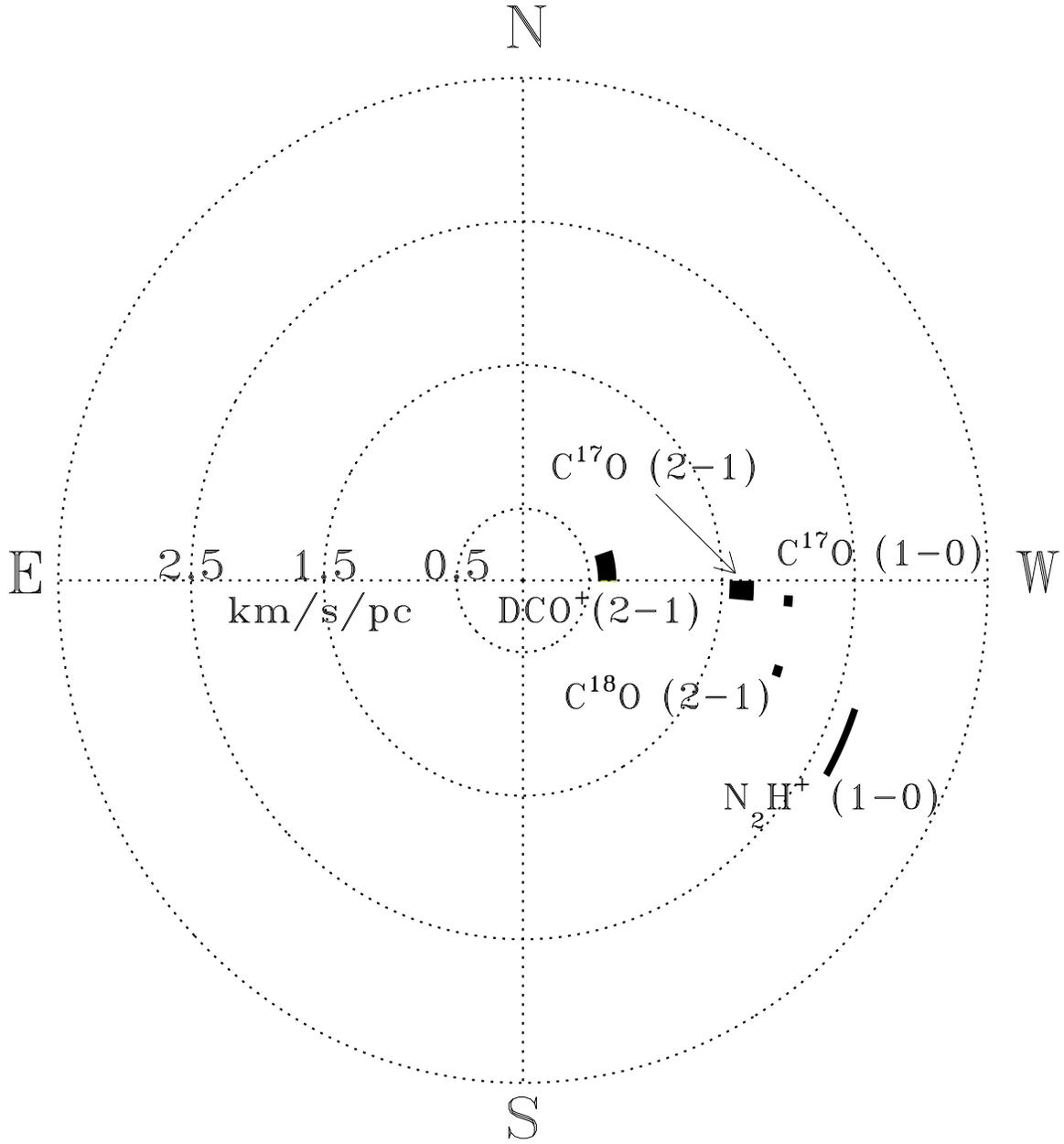}
\caption{The overall velocity gradient for each tracer in the region
  bounded by $-140 < $ RA$_{\rm offset} < 70$ and $-20 < $ DEC$_{\rm
  offset} < 70$, the region in the red box in Fig.~\ref{COARROWS} and
  \ref{COARROWS}.  The dotted circles show the magnitude of
  gradients equal to 0.5, 1.5, 2.5 and 3.5 km s$^{-1}$ pc$^{-1}$.
  \label{CENTERGRAD}}
\end{figure}

\begin{figure}
\epsscale{0.7}
\plotone{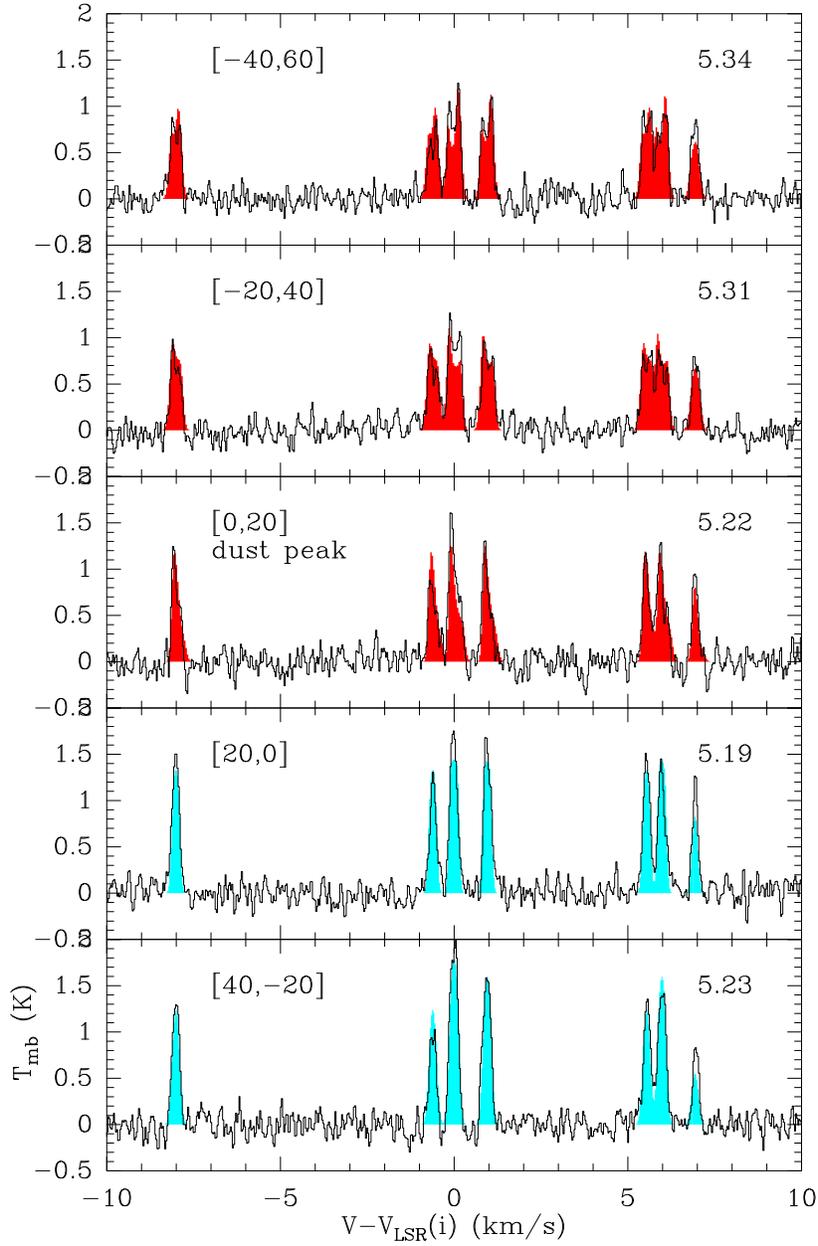}
\caption{\NTHP (1--0) spectra (black histogram) along a South--East
  (bottom panel) to North--West (top panel) strip (see spectra
  labelled with colored dots in Fig.~\ref{spectra_map}).  Filled red
  histograms are two--layer model results, whereas light blue filled
  histograms are simply fits to the line, assuming one emitting layer
  and constant $T_{\rm ex}$ for the seven hyperfines.  Offsets from
  Fig.~\ref{spectra_map} are in the top left of each panel.  Numbers
  in the top right are the $V_{\rm LSR}$ velocities (in \kms ) from gaussian fits
  to the hfs structure. The parameters of the fits are in
  Table~\ref{table_layer}. Note that the redshifted absorption (and thus 
the infall velocity) is largest
  toward the dust peak and that the self-absorption is present towards
  the North--West but not towards the South--East.  \label{two_layer}}
\end{figure}

In Fig.~\ref{two_layer} we present five spectra which represent a cut
across the major axis of the core, passing through the dust peak. For
display purposes, the spectra have been centered to 0 velocity,
subtracting the LSR velocity obtained from a Gaussian fit to the weak
hyperfine component (for offsets [-40,60], [-20,40], [0,20], where the
self-absorption in present) or hfs fits in CLASS\footnote{CLASS is
part of the Grenoble Image and Line Data Analysis Software (GILDAS),
available at http://www.iram.fr/IRAMFR/GILDAS/gildas.html).} (for
offsets [20,0] and [40,-20]). The $V_{\rm LSR}$ velocity is shown in
the top right of each panel.  The cut is from South-East (offset
[40,-20], see Fig.~\ref{spectra_map}) to North-West (offset
[-40,60]). The first thing to note in the figure is that clear signs
of self-absorption and asymmetry are present toward the dust peak and
in the North-West, but not in the two Southern positions.  This trend
can also be seen as a general feature in Fig.~\ref{spectra_map}, where
it is evident that asymmetric lines are more numerous North--West of
the dust peak.

The excitation temperature, total optical depth, line width and the
velocity ($V-V_{\rm LSR}$, see Fig.~\ref{two_layer}) of the foreground
(F) and background (B) layers are reported in Table \ref{table_layer}.
To find the best fit parameters, we first performed an hfs fit to the
[20,0] spectrum, which is the closest spectrum to the dust peak not
showing self-absorption.  The values of $\tau_{\rm TOT}$, $T_{\rm ex}$
and line width obtained from this fit have been adopted for the
background emission at the dust peak position and the best fit has
been found by adding the foreground layer and minimising the
residuals. For the two spectra North--West of the [0,20] position,
adjustment to the parameters of the background layer were necessary to
obtain a good fit.  We point out that the five spectra we chose for
this analysis are representative of the whole area surrounding the
TMC--1C dust peak, where a mixture of symmetric, blue--shifted and
red--shifted spectra are present.  As already stated, the majority of
the asymmetric spectra show inward motions and extend over a region
with radius $\sim$7000 AU (see Fig.~\ref{spectra_map}).

The properties derived for the foreground layer ($T_{\rm ex}$
$\simeq$3.3--3.5~K, $\tau_{\rm TOT}$ $\simeq$10--15, and $\Delta v
\simeq$0.2 \kms) have been used as input parameters in a Large
Velocity Gradient (LVG) code\footnote{available at
http://www.strw.leidenuniv.nl/$\sim$moldata/radex.php} for a uniform medium
and found to be consistent with the \NTHP (1--0) tracing gas at a
density $n(\MOLH )$ $\simeq$ 5$\times$10$^3$ \percc , kinetic
temperature $T_{\rm kin}$ $\simeq$ 10~K and with column density
$N(\NTHP )$ $\simeq$ 5$\times$10$^{12}$ \cmsq , values comparable to
those found for the background layer (see Table \ref{FDTABLE} and
Sec.~\ref{GASCOLUMN}).

It is interesting that the maximum of the line-of-sight component of
the inward velocity ($\sim$0.15 \kms ) is found toward the dust peak,
whereas one pixel away from it, the inward velocity drops to 0.05
$\kms $. This is suggestive of a geometric effect, in which the inward
velocity vector is directed toward the dust peak, so that only a
fraction $cos (\theta)$ (with $\theta$ the angle between the
l.o.s. and the infall velocity direction) of the total velocity is
directed along the line of sight in those positions away from the dust
peak.  Of course, our simplistic model prevents us to go further than
this, i.e. the uncertainties are too large to build a 3D model of the
velocity profile within the cloud.

As shown in Fig.~\ref{spectra_map}, in the North--West end of the
TMC--1C core (around offset [-250,150]), there are other signatures of
inward motions, which may indicate the presence of another
gravitational potential well.  This suggestion is indeed reinforced by
Fig.~8 of \citet{Schnee06b}, which shows high extinction and low
temperatures in the same direction. Unfortunately, the continuum
coverage is not good enough to attempt a detailed analysis, but it
appears evident that the extension toward the North--West is another
dense core connected to the main TMC--1C condensation with lower
density and warmer gas and dust.

\section{Analysis. II. Column Density and Temperature} \label{ANALYSIS_II}

\subsection{Gas Column Density} \label{GASCOLUMN}

To derive the column density of gas from each molecule, we assume that
all rotation levels are characterized by the same excitation
temperature $T_{ex}$ (the CTEX method, described in Caselli et
al. 2002b).  In case of optically thin emission,
\begin{equation} \label{NTOT}
N_{tot} = \frac{8 \pi W}{\lambda^3 A} \frac{g_l}{g_u}
          \frac{1}{J_{\nu}(T_{ex}) - J_{\nu}(T_{bg})}
          \frac{1}{1 - \exp(-h \nu/k T_{ex})}
          \frac{Q_{rot}}{g_l \exp(-E_l/k T_{ex})}
\end{equation}
where $\lambda$ and $\nu$ are the wavelength and frequency of the
transition, $k$ is the Boltzmann constant, $h$ is the Planck constant,
$A$ is the Einstein coefficient, $g_l$ and $g_u$ are the statistical
weights of the lower and upper levels, $J_{\nu}(T_{ex})$ and
$J_{\nu}(T_{bg})$ are the equivalent Rayleigh-Jeans excitation and
background temperatures, W is the integrated intensity of the line.
The partition function ($Q_{rot}$) and the energy of the lower level
($E_l$) for linear molecules are given by:
\begin{equation} \label{QROT}
Q_{rot} = \sum_{J=0}^{\infty} (2J + 1)\exp(-E_J/kT)
\end{equation} 
and
\begin{equation} \label{EJ}
E_J = J(J+1)hB
\end{equation}
and $B$ is the rotational constant (see Table \ref{CONSTANTS5} for the
values of the constants). 

The \CSEO(1--0) and (2--1) lines have hyperfine structure, enabling
the measurement of the optical depth. We find that the lines are
optically thin throughout the core.  To determine the column density,
we assume an excitation temperature of 11~K, which is the average
value of $T_{\rm ex}$ found from our \CEIO \ data around the dust peak
position, as explained in Sect.~\ref{GASTEMP}.  In the case of \CEIO
(2--1) lines, we correct for optical depth before determining the
column density, using the correction factor:
\begin{eqnarray} \label{CTAU}
C_{\tau} & = & \frac{\tau}{1-e^{-{\tau}}}
\end{eqnarray}

As explained in Sect.~\ref{RESULTS_SPECTRA}, \NTHP (1--0) lines show
clear signs of self--absorption in an extended area around the dust
peak.  To determine the column density across the core, first we
select spectra without self--absorption, and those with high S/N
(i.e. with $W/\sigma_W$ $>$ 20, with $W$ $\equiv$ integrated
intensity; see Caselli et al. 2002b) have been fitted in CLASS to find
$T_{\rm ex}$ and $\tau$.  The mean value of $T_{\rm ex}$ found with
this analysis (4.4$\pm$0.1~K) has been used for all other positions
where an independent estimate of $T_{\rm ex}$ was not possible
(i.e. for self-absorbed or thin lines).  For optically thin \NTHP
(1--0) transitions, the intensity was integrated below the seven
hyperfine and the expression (\ref{NTOT}) used to determine the total
column density. 

In cases of self-absorbed spectra, the \NTHP \ column density has been
estimated from the integrated intensity of the weakest (and lowest
frequency) hyperfine component ($F_1 F$ = 1~0$\rightarrow$1~1), using
eq.~\ref{NTOT} (assuming $T_{\rm ex}$=4.4~K) and multiplying by 27
(the inverse of the hyperfine relative intensity).  The weakest
component is not affected by self--absorption, as shown in
Fig.~\ref{infall_dust_peak}, suggesting that its optical depth is low.
We have checked that these two different methods approximately give
the same results by measuring the \NTHP\ column density with both
procedures in those cases where self-absorption is not present and
where the weakest hyperfine component has a S/N ratio of at least 4.
We found that the two column density values agree to within 10\%.

~\DCOP (2--1) and \DCOP (3--2) lines are clearly self--absorbed and
the column density determination is very uncertain (given that there
are no clues about their optical depth and excitation
temperature). The estimates listed in Table~\ref{FDTABLE} should be
considered lower limits. $T_{\rm ex}$ = 4.4~K has beed assumed, based
on the fact that the \DCOP \ lines are expected to trace similar
conditions than \NTHP.  C$^{34}$S(2--1) spectra have low sensitivity,
and the lines are affected by self--absorption (see
Fig.~\ref{IRAMSPEC}), so the derived C$^{34}$S column density is
highly uncertain.

\begin{figure}
\epsscale{1.0}
\plotone{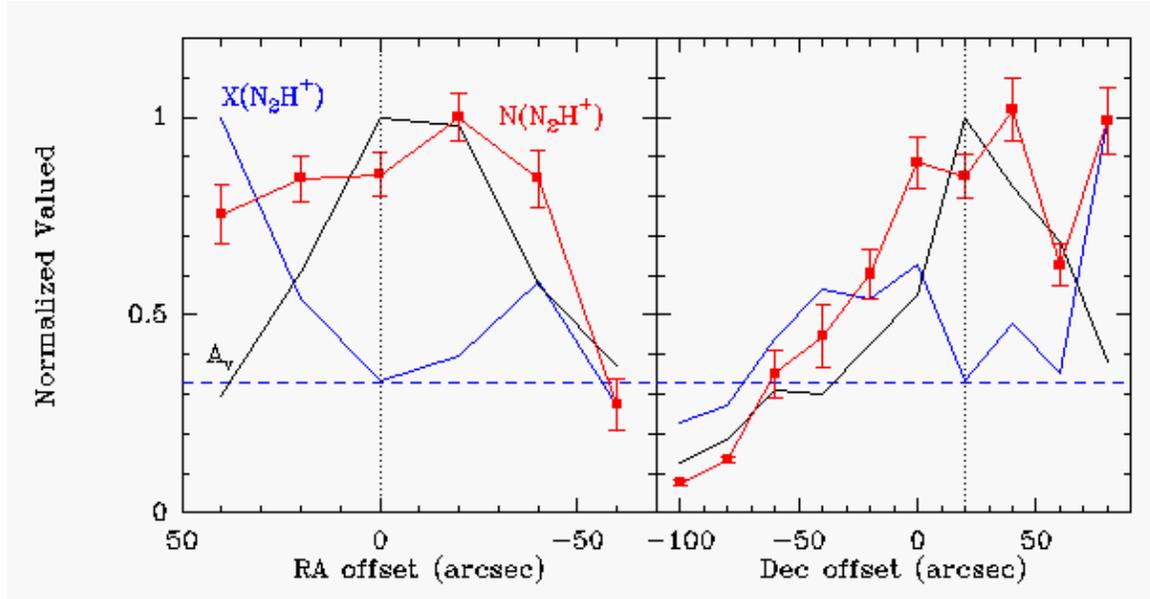}
\caption{Right ascension (left) and declination (right) cuts 
  cross the TMC--1C core, passing through the dust peak position
  (black dotted line), of $A_{\rm V}$ (black curve), $N(\NTHP )$ (red
  curve), and $X(\NTHP )$ (blue curve) normalized to their maximum
  values (see text). The blue dashed line is the abundance value
  observed across L1544 (see Vastel et al. 2006).  $X(\NTHP )$ shows
an anticorrelation with the dust profile, suggestive of some \NTHP\
depletion toward the dust center. \label{CUT}}
\end{figure}

The \NTHP \ abundance, $X(\NTHP)$ ($\equiv$ $N(\NTHP)/N(\MOLH )$),
toward the dust peak is 1.6$\times$10$^{-10}$, identical (within the
errors) to that derived toward the L1544 dust peak
\citep[e.g.]{Crapsi05}.  This is interesting considering that
$N(\MOLH)$ in TMC--1C is 1.6 times lower than in L1544, in which
\NTHP\ closely follows the dust column (as already found in previous
work).  However, unlike L1544, where the \NTHP\ abundance appears
constant with impact parameters (e.g. Tafalla et al. 2002 and
Vastel et al. 2006), in TMC--1C the \NTHP\ abundance increases away
from the dust peak by a factor of about two within 50\arcsec, as shown
in Fig.~\ref{CUT} (see also Fig.~\ref{IRAMFDNH} in Sect.~\ref{smol}).
Fig.~\ref{CUT} displays $A_{\rm V}$ (see Sect.~\ref{DUSTCOLUMN}),
$N(\NTHP)$ and $X(\NTHP)$, normalized to the corresponding maximum
values (63.2 mag, 1.1$\times$10$^{13}$ \cmsq , and
4.8$\times$10$^{-10}$, respectively) in two cuts (one in right
ascension and one in declination) passing through the dust peak.  One
point to note is that the abundance derived at the dust peak (marked
by the black dotted line) is the {\it minimum} value observed,
indicating moderate (factor of $\sim$2) depletion.

\subsection{Dust Column Density} \label{DUSTCOLUMN}

In \citet{Schnee06b} we used SCUBA and MAMBO maps at 450, 850 and 1200
\micron\ to create column density and dust temperature maps of TMC--1C.
In this paper we smooth the dust continuum emission maps to the
20\arcsec\ spacing of the IRAM maps and then derive $A_{\rm V}$ and
$T_d$ to facilitate a direct comparison of the gas and dust
properties.  At each position, we make a non-linear least squares fit
for the dust temperature and column density such that the difference
between the predicted and observed 450, 850 and 1200 \micron\
observations is minimized.  The errors associated with such a fitting
procedure are described in \citet{Schnee06b}.  Dust column density and
temperature maps of TMC--1C are shown in Fig.~8 of \citet{Schnee06b}.

\begin{figure}
\epsscale{1.0}
\plotone{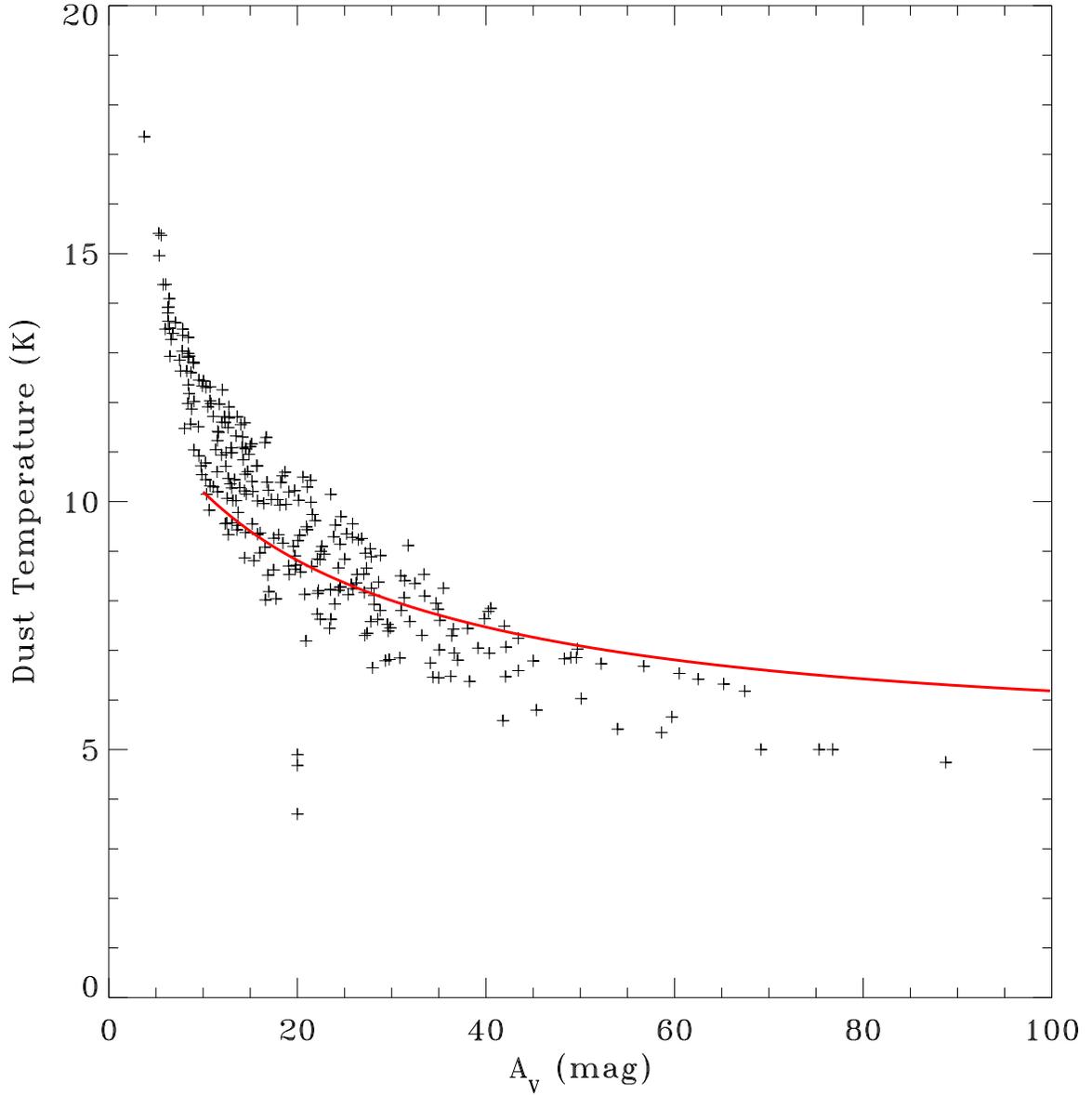}
\caption{The dust temperature plotted against the column density,
  derived from the dust emission maps at 450, 850 and 1200 \micron.
  The spatial resolution of the data is 14\arcsec, resolving the
  innermost region of the core, with $A_{\rm V}$ $\simeq$ 90~mag and
  $T_{\rm dust}$ $\simeq$ 5~K.  The solid red line shows the $T_d-A_V$
  relation predicted for an externally heated pre-protostellar core by
  \citet[Eq. 26]{Zucconi01}. Departures from spherical symmetry are 
probably causing the observed discrepancy between observations 
and (spherically symmetric) model predictions.
         \label{TDVSAV}}
\end{figure}

From Fig.~8 in \citet{Schnee06b}, it is clear that there is an
anti-correlation between extinction and dust temperature (as also
predicted by theory, e.g. Evans et al. 2001, Zucconi et al. 2001,
Galli et al. 2002). To better show this, the two quantities are
plotted in Fig.~\ref{TDVSAV}. The data in Fig.~\ref{TDVSAV} are not
smoothed to the IRAM 30m beam at 3~mm, since this is only a dust
property intercomparison and does not refer to the gas properties.
Higher $A_{\rm V}$ (60 $<A_{\rm V}\leq$ 90~mag) and lower dust
temperatures (5 $\leq T_{\rm dust} <$ 6~K) are detectable at the
higher resolution.  We compare the $T_{\rm d}-A_{\rm V}$ relationship
seen in TMC--1C with that predicted by \citet{Zucconi01} for an
externally heated pre-protostellar core (the solid red line in
Fig.~\ref{TDVSAV}).  We find that at high column density ($A_{\rm V} >
30$) the observed dust temperature in TMC--1C is lower than that of
the model core, while at low column density ($10 < A_{\rm V} < 20$)
the observed dust temperature is higher than the model predicts.
However, given that the model predicts the dust temperature at the
center of a {\it spherical} cloud, and that the geometry of TMC--1C is
certainly not spherical, only a rough agreement between the model and
observations should be expected.

\subsection{Gas Temperature} \label{GASTEMP}

Because of its low dipole moment, CO is a good gas thermometer, given
that it is easily thermalized at typical core densities. However, it
is now well established that CO is significantly frozen onto dust
grains at densities $\ge$ 10$^5$ cm$^{-3}$ (one exception being
L1521E; Tafalla \& Santiago, 2004b) and this is also the case in
TMC--1C. Therefore, at the dust peak we do not expect to measure a gas
temperature from CO of $\sim$7~K, but instead a higher value
reflecting the temperature in the outer layers of the cloud.  The
lines available for this analysis are: C$^{17}$O(1--0),
C$^{17}$O(2--1), and C$^{18}$O(2--1).  The
C$^{17}$O(2--1)/C$^{18}$O(2--1) brightness temperature ratio has been
used to derive the excitation temperature of the C$^{18}$O line, which
is coincident with the kinetic temperature if the line is
thermalized. To test the hyposthesis of thermalization we use an LVG
(Large-Velocity-Gradient) program to determine at which volume density
and kinetic temperature the observed C$^{17}$O(1--0) and
C$^{17}$O(2--1) brightness temperatures can be reproduced.  We use a
one-dimensional non-LTE radiative transfer code \citep{vanderTak07}
available at http://www.strw.leidenuniv.nl/~moldata/radex.html.

\subsubsection{C$^{17}$O(2-1) and C$^{18}$O(2-1) 
as a measure of $T_{\rm kin}$}

These two lines have similar frequencies, so the corresponding angular
resolution is almost identical and no convolution is needed.
Following a similar analysis done with the J=1--0 transition of the
two CO isotopologues (Myers et al. 1983), the optical depth of the
C$^{18}$O(2--1) line ($\tau_{18}$) can be found from:
\begin{eqnarray}
\frac{T_{\rm mb}[\CEIO (2-1)]}{T_{\rm mb}[\CSEO (2-1)]} & = & 
3.65 \times \frac{1 - e^{-\tau_{18}}}{\tau_{18}} ,
\end{eqnarray}
where $T_{\rm mb}[i]$ is the main beam brightness temperature of
transition $i$ (assuming a unity filling factor).  The last term in
the right hand side is the optical depth correction which is used to
determine the total column density of \CEIO \ in a plane parallel
geometry, which most likely applies to CO emitting regions, i.e. the 
external core layers.

Once $\tau_{18}$ is measured, the excitation temperature ($T_{\rm
ex}$) of the corresponding transition (thus the gas kinetic
temperature, if the line is in local thermodynamic equilibrium) can be
estimated from the radiative transfer equation:
\begin{eqnarray}
T_{\rm mb} & = & [ J_{\nu }(T_{\rm ex}) - J_{\nu}(T_{\rm bg}) ] 
(1 - e^{-\tau}) ,
\label{erad}
\end{eqnarray}
where $ J_{\nu }(T_{\rm ex})$ and $J_{\nu}(T_{\rm bg})$ are the 
equivalent Reyleigh--Jeans temperatures, with 
\begin{eqnarray}
J_{\nu}(T) = \frac{T_0}{\exp(T_0/T)-1} ,
\end{eqnarray}
$T_0$ = $h\nu /k_{\rm B}$, and $\nu$  the frequency of the \CEIO (2--1) line
(see Table \ref{IRAMOBS}).

Figure \ref{ftemp} (left panel) shows the results of this analysis.
The set of data points in Fig.~\ref{ftemp} is limited to only those
spectra with $T_{\rm mb}[\CSEO (2--1)]/T_{\rm rms}$ $\ge$ 10, to avoid
scatter due to noise. The error associated with the gas temperature
has been calculated by propagating the errors on $\tau_{18}$ and
$T_{\rm mb}$[\CEIO (2--1)] into Eq.~\ref{erad} and its expression is
given in the appendix.

It is interesting to note that the \CEIO (2--1) excitation temperature
appears to decrease away from the center, and, if this line is
thermalized, it suggests that the kinetic temperature also drops away
from the center, in contrast with the dust temperature. Indeed, the
two quantities are completely uncorrelated in the 12 common positions
with large S/N \CEIO (2--1) spectra (not shown).  $T_{\rm ex}$ is close
to 12~K at the core center, whereas it drops to 9--10~K one arcmin
away from the dust peak.  Is this drop due to a decreasing gas
temperature, as recently found by \citet{Bergin06} in the Bok Globule
B68?  Unlike B68, we believe that our result is due to the volume
density decrease.  Indeed, the critical densities of the J = 2--1
lines of \CSEO \ and \CEIO \ are a few $\times$ 10$^{4}$ \percc , so
that only if the volume density traced by one of the two isotopologues
is larger than, say, 5$\times$10$^4$ \percc \ can the J = 2--1 lines
be considered good gas thermometers.  In the next subsection, we
investigate this point more quantitatively.

\subsubsection{C$^{17}$O(1--0) and C$^{17}$O(2--1) to measure 
$T_{\rm ex}$[\CSEO (2--1)]}   

\begin{figure}
\epsscale{1.0}
\plottwo{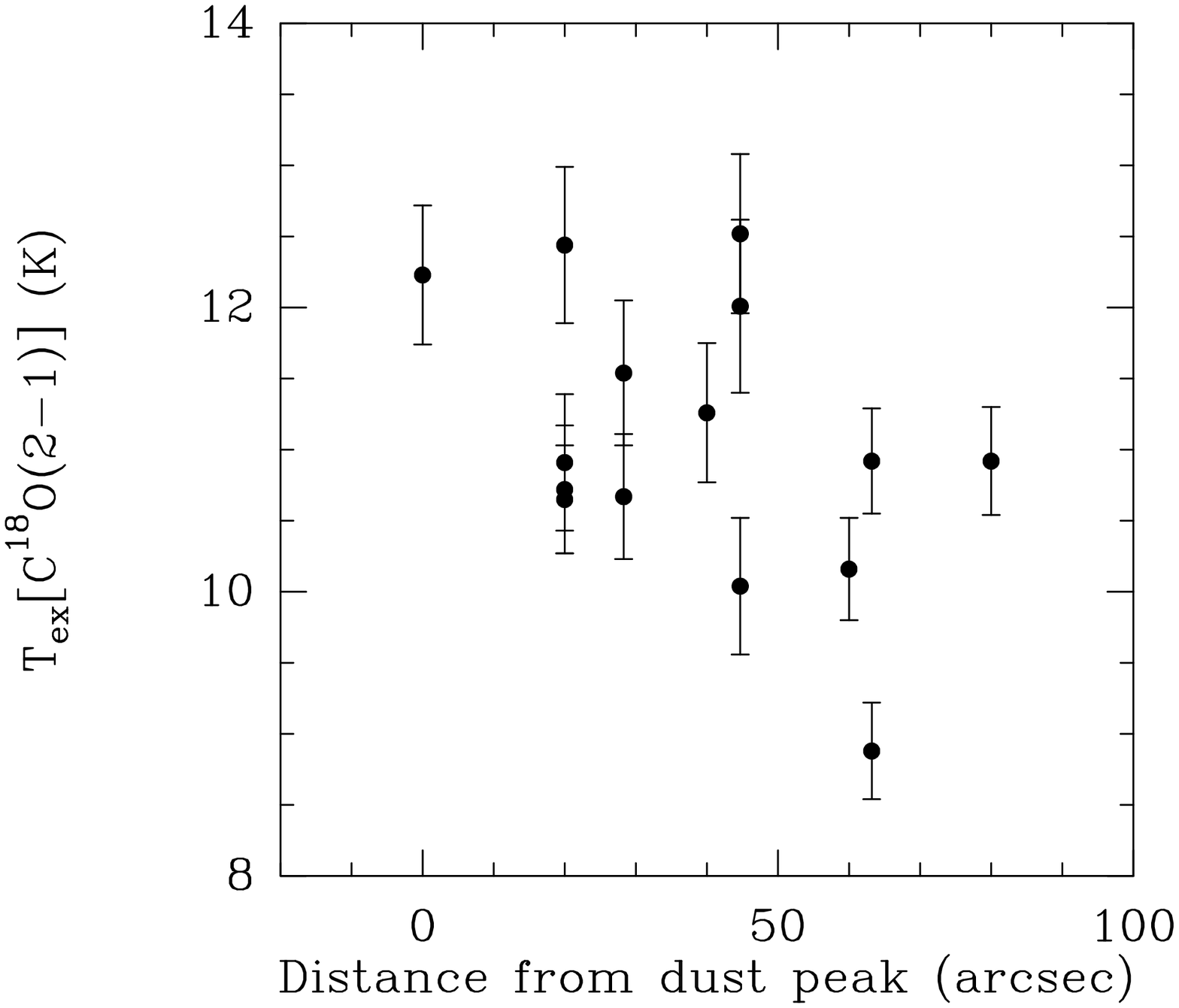}{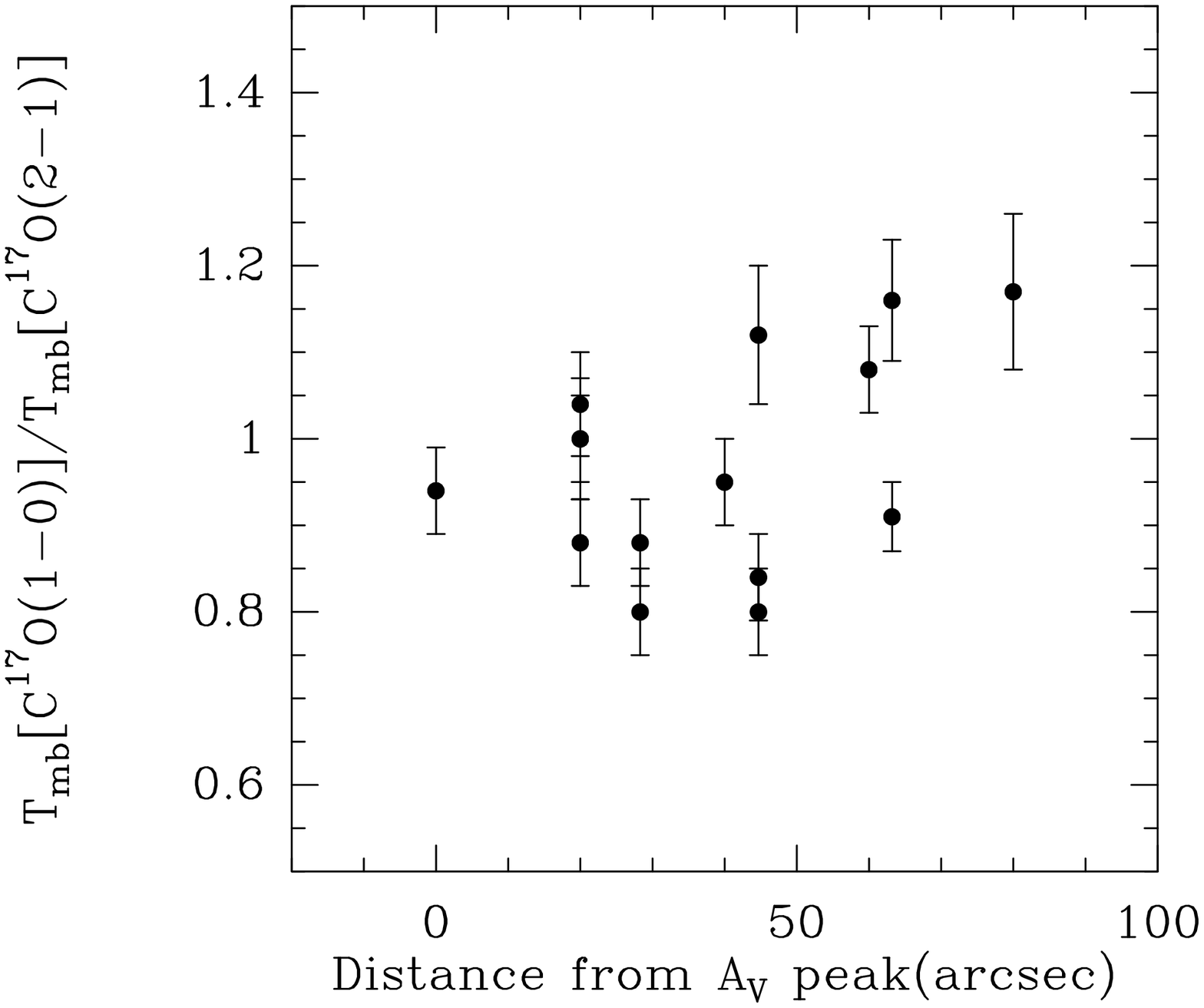}
\caption{({\it left}) Excitation temperature of the \CEIO (2--1) line. 
  If this line is in LTE, the present data suggest a gas temperature
  decrease from about 12~K at the dust peak, to about 10~K at $\ge$
  1\arcmin \ away from the dust peak (equivalent to a projected
  distance of $\ge$8,000 AU).  But we believe that this is caused by
  the density drop traced by the \CEIO (2--1) line (see text for
  details). ({\it right}) Brightness temperature ratio of the \CSEO
  (1--0) and \CSEO (2--1) lines as a function of distance from the
  dust (or $A_{\rm V}$) peak.  The ratio increases with distance,
  suggesting that some physical property ($T_{\rm kin}$ and/or
  $n(\MOLH )$) changes as well.
\label{ftemp}}
\end{figure}

In Fig.~\ref{ftemp} (right panel), the brightness temperature ratio of
the \CSEO (1--0) and \CSEO (2--1) lines is plotted as a function of
distance from the dust peak.  Because of the different angular
resolutions at the 2--1 and 1--0 frequencies, the 1 mm data have been
smoothed to the 3 mm resolution and both data cubes have then been
regridded, to allow a proper comparison.  The ratio is indeed
increasing towards the edge of the cloud, consistent with our previous
finding of a $T_{\rm ex}$[\CEIO (2--1)] drop in the same direction
(see left panel).

Both \CSEO (1--0) and (2--1) lines possess hyperfine structure, which
provides a direct estimate of the line optical depth.  Using the hfs
fit procedure available in CLASS, we found that all over the TMC--1C
cloud, both lines are optically thin.  This means that it is not
possible to derive the excitation temperature in an analytic way, so
we use the LVG code introduced in Sect.~\ref{GASTEMP}.  This code
assumes homogeneous conditions, which is likely to be a good
approximation for the region traced by CO isotopologues. In fact,
because of freeze-out, CO does not trace the regions with densities
larger than about 10$^5$ cm$^{-3}$ (see below and Sect.~\ref{smol}),
so that the physical conditions traced by CO around the dust peak are
likely to be close to uniform (n(H$_2$) $\sim$ a few times 10$^4$ and
about constant temperature).  This is also supported by the integrated
intensity CO maps, which appear extended and uniform around the dust
peak (see Fig.~\ref{INTMAPS}).

\begin{figure}
\epsscale{0.7}
\plotone{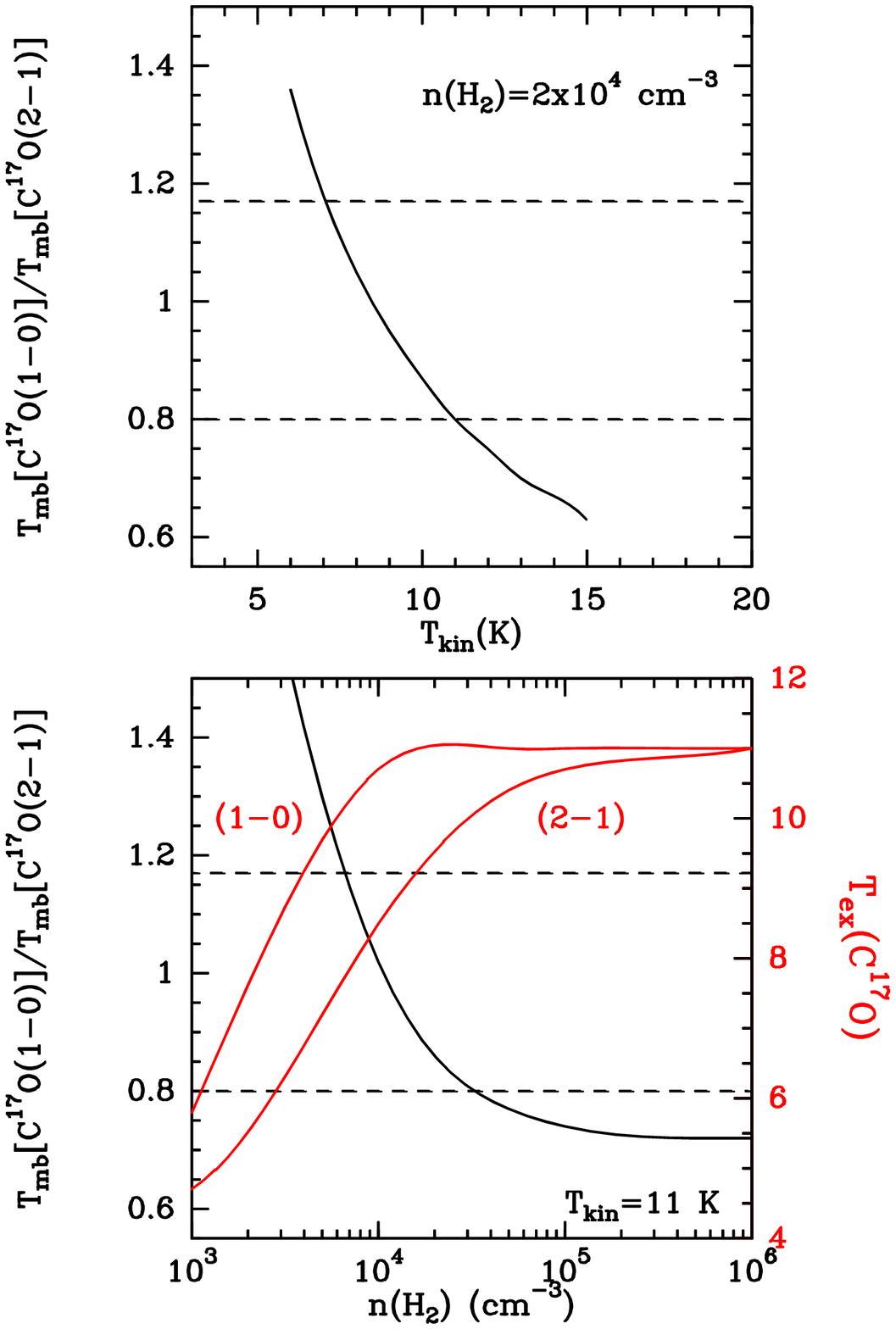}
\caption{Ratio of the \CSEO (1--0) and \CSEO (2--1) brightness
  temperatures as a function of ({\it Top}) gas temperature, for a
  fixed value of the volume density ($n({\rm \MOLH})$ =
  2$\times$10$^4$ \percc), and ({\it Bottom}) volume density, for a
  fixed value of the gas temperature ($T_{\rm kin}$ = 11~K). In both
  panels, $N(\CSEO )$ = 10$^{15}$ \cmsq \ and $\Delta v$ = 0.4~\kms
  . The dashed horizontal lines enclose the observed range of $T_{\rm
  mb}$ ratios (see Fig.~\ref{ftemp}, right panel).  Note that
  observations are both consistent with a gas temperature {\it
  decrease} (and high constant density), as well as with a volume
  density decrease (and constant gas temperature) away from the dust
  peak. We believe that the latter is a more plausible solution than
  the former. \label{flvg}}
\end{figure}

To better understand this result, the LVG code has been run to see how
changes in volume density and gas temperature affect the line ratio.
This is shown in Fig.~\ref{flvg}, where the top panel shows the
$T_{\rm mb}$ ratio as a function of $T_{\rm kin}$ for a fixed value of
the volume density ($n(\MOLH )$ = 2$\times$10$^4$ \percc ), a \CSEO \
column density of 10$^{15}$ \cmsq \ (as found in Section
\ref{GASCOLUMN}), and a line width of 0.4\kms , as observed. The 
horizontal dashed lines enclose the range of $T_{\rm mb}$ ratios
observed in TMC--1C and reported in Fig.~\ref{ftemp}.  Thus, the
observed $T_{\rm mb}$ range (at this volume density) corresponds to a
range of gas temperature between 11~K toward the dust peak and
$\sim$7~K away from it, thus confirming our previous findings of a
decreasing \CEIO (2--1) excitation temperature away from the dust
peak.

In the bottom panel of Fig.~\ref{flvg}, the same brightness
temperature ratio is plotted as a function of $n(\MOLH )$, for a fixed
kinetic temperature ($T_{\rm kin}$=11~K), $N(\CSEO )$ = 10$^{15}$
\cmsq\ and $\Delta v$=0.4~\kms , as before. The black curve shows this 
variation and, not surprisingly, the observed range of $T_{\rm mb}$
ratios can also be explained if the volume density (traced by the
\CSEO\ lines) decreases from $\simeq 4\times10^4$ \percc\ toward the
dust peak to $\simeq 6\times 10^3$ \percc\ away from it (the point
farthest away being at a projected distance of 80\arcsec , or
11,000~AU, see Fig.~\ref{ftemp}).  Note that the volume density traced
by the \CSEO\ line towards the dust peak is significantly lower than
the central density of TMC-1C ($\simeq$5$\times$10$^5$ \percc ; see
Schnee et al. 2007), once again demonstrating that CO is not a good
tracer of dense cores.  The bottom panel of Fig.~\ref{ftemp} is
consistent with a volume density {\it decrease} away from the dust
peak, or, more precisely, a lower fraction of (relatively) dense gas
intercepted by the \CSEO\ lines along the line of sights.  In the same
plot, the red curves show the excitation temperatures of the \CSEO
(1--0) and (2--1) lines vs. $n(\MOLH )$.  Note that $T_{\rm ex}$[\CSEO
(2--1)] = $T_{\rm kin}$ only when the density becomes larger than
$\sim$10$^5$ \percc . Thus, the \CSEO(2--1) line is sub-thermally
excited in TMC--1C.

In summary, the rise in the \CSEO (1--0)/(2--1) brigthness temperature
ratio away from the dust peak ratio can be caused by {\it either} a
gas temperature decrease {\it or} a volume density decrease (or both).
Considering that the dust (and likely the gas; see the recent paper by
Crapsi et al. 2007) temperature is clearly increasing away from the
dust peak, we believe that the drop in $T_{\rm ex}$ observed both
using the \CEIO \ and \CSEO \ lines is more likely due to a drop in
the volume density traced by these species.  This is reasonable in the
case of a core embedded in a molecular cloud complex, such as TMC--1C,
where the fraction of low density material intercepted along the line
of sight by \CEIO \ and \CSEO \ observations is significantly larger
than in isolated Bok Globules such as B68 (see Bergin et al. 2006).
In any case, a detailed study of the volume density structure of the
outer layers of dense cores will definitely help in assessing this
point.

\section{Analysis. III. Chemical Processes} \label{ANALYSIS_III}

\subsection{Molecular Depletion}
\label{smol}

By comparing the integrated intensity maps of CO isotopologues and
\nthp (1--0) (Fig.~\ref{INTMAPS}) in TMC-1C with the column density
implied by dust emission (Fig.~8 in \citet{Schnee06b}), we see that at
the location of the dust column density peak the CO emission is not
peaked at all.  The \nthp (1--0) emission peaks in a ridge around the
dust column density maximum, not at the peak, but in general \nthp\
traces the dust better than the \ceo\ emission does.  Below, we
measure the depletion of each observed molecule and compare our
results to similar cores.

Previous molecular line observations of starless cores such as L1512,
L1544, L1498 and L1517B, consistently show that CO and its
isotopologues are significantly depleted, e.g. \citep{Lee03,
Tafalla04}.  However, other molecules such as \dcop\ and \nthp\ are
typically found to trace the dust emission well, (e.g. Caselli et
al. 2002b; Tafalla et al. 2002, 2004), although there is some evidence
of their depletion in the center of chemically evolved cores, such as
B68 \citep{Bergin02}, L1544 \citep{Caselli02c} and L1512
\citep{Lee03}.  In order to measure the depletion in TMC-1C, we define
the depletion factor of species $i$:
\begin{equation} \label{FD}
f_D(i) = X_{can}(i) \frac{N(H_2)_{dust}}{N(i)}
\end{equation}
where $X_{can}$ is the ``canonical'' (or undepleted) fraction
abundance of species $i$ with respect to H$_2$ (see Table
\ref{CONSTANTS5}), $N(H_2)_{dust}$ is the column density of molecular
hydrogen as derived from dust emission, and $N(i)$ is the column
density of the molecular species as derived in Section
\ref{GASCOLUMN}.  

\begin{figure}
\epsscale{0.8}
\plotone{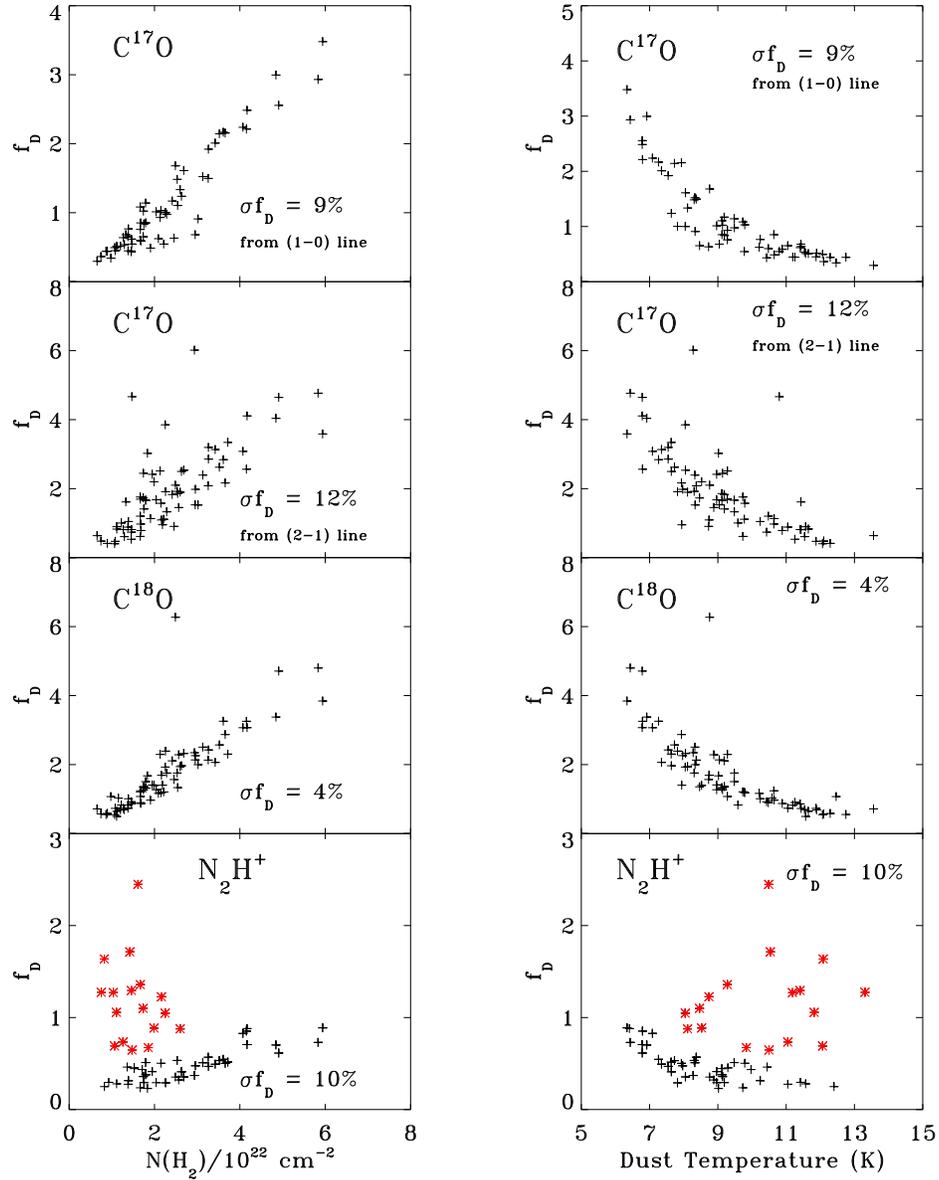}
\caption{({\it left}) The depletion factor derived from molecular
  transitions observed with the IRAM 30m plotted against the column
  density derived from dust emission maps at 450, 850 and 1200
  \micron.  The median percent error in the depletion (calculated from
  the RMS of the spectrum) is given in each panel.  ({\it right}) Same
  as left, except that the depletion factor is plotted against the
  dust temperature.  Those points with anomalously high \NTHP depletion
  factors ($N_{H_2} < 3\times10^{22}$ and $f_D > 0.6$) are shown in red, and
  their positions are shown with red crosses in Fig.~\ref{INTMAPS}.  The \NTHP\
depletion factor goes below 1 because of our arbitrary choice of the \NTHP\ 
column density (see text). 
  \label{IRAMFDNH}}
\end{figure}

The derived depletion factors for each molecule (except for \DCOP \
and C$^{34}$S, where the column density determination is quite
uncertain, as explained in Sec.~\ref{GASCOLUMN}) in each position with
signal to noise $>$ 3 are plotted against the dust-derived column
density in Figure \ref{IRAMFDNH}.  The typical random error in the
derived depletion is shown in each panel, and is derived from the
noise in the spectra.  Uncertainty in the derived column density from
dust emission is dominated by calibration uncertainties in the
bolometer maps, and is not included in this calculation, nor is the
uncertainty in the calibration of the spectra ($\sim$20\%), which
would adjust the derived depletion factors systematically, but would
not alter the observed trends.  Depletions factors are found to
increase with higher dust colunm density (Fig.~\ref{IRAMFDNH}).  In
TMC-1C we see a linear relationship between \cso\ and \ceo\ depletion
and dust-derived column density, which has also been seen in \ceo\ by
\citet{Crapsi04} in the core L1521F, which contains a Very Low Luminosity 
Object \citep{Bourke06}.  To check the impact of resolution on the
derived depletion, we compare the \ceo (2--1) depletion when smoothed
to 20\arcsec\ (the resolution of the \cso (1--0) data) with that
derived from smoothing \ceo (2--1) to 14\arcsec\ (the resolution of
the bolometer data).  We find no systematic difference between the two
calculations of the depletion, and a 13\% standard deviation in the
ratio of the derived depletions.

The depletion factor and column density at the position of the dust
peak is listed in Table \ref{FDTABLE} for each tracer.  The depletion
factor measured in \NTHP\ clearly follows a different trend compared
to the CO isotopologues.  First of all, in Fig.~\ref{IRAMFDNH} the
\NTHP\ depletion factor is allowed to have values below 1, because of
our (arbitrary) choice for the ``canonical'' abundance of \NTHP\,
assumed here to be equal to 1.4$\times 10^{-10}$, the average value
across TMC--1C. We point out that a ``canonical'' abundance for \NTHP\
is much harder to derive than for CO, because \NTHP\ lines are much
harder to excite (and thus detect) in low density regions where
depletion is negligible. Nevertheless, Fig. 20 shows that the \NTHP\
depletion factor monotonically increases (as in the case of CO) for
N(H$_2$) $\geq 3\times 10^{22}$ cm$^{-2}$ (or A$_{\rm V} \geq 30$ mag).
{\it This is clear evidence of \NTHP\ depletion in the core nuclei, in
a central region with radius $\sim$6000 AU, where A$_{\rm V} \geq 30$
mag} (see also Fig.~\ref{CUT}).

The dispersion in the \NTHP\ depletion factor vs. N(H$_2$) relation is
very large with no obvious trend at lower A$_{\rm V}$ values (N(H$_2$)
$< 3\times 10^{22}$ cm$^{-2}$), which we believe is due to our choice
of excitation temperature where the \NTHP (1--0) line is optically
thin.  As explained in Sect.~\ref{GASCOLUMN}, in the case of optically
thin lines, T$_{\rm ex}$ has been assumed equal to 4.4 K, the mean
value derived from the optically thick spectra which do not show
self-absorption (and which trace regions with A$_{\rm V} \geq 20$
mag).  Therefore, in all positions with N(H$_2$) below 2$\times
10^{22}$ cm$^{-2}$, where the volume density is also likely to be low,
the assumed \NTHP (1--0) excitation temperature is likely to be an
overestimate of the real T$_{\rm ex}$.  To see if this can indeed be
the cause of the observed scatter, consider a cloud with kinetic
temperature of 10 K, volume density of 3$\times 10^4$ cm$^{-3}$, line
width of 0.3 km s$^{-1}$ and an excitation temperature of 4.4 K for
the \NTHP (1--0) line.  Using the RADEX LVG program, this corresponds
to a \NTHP\ column density of 10$^{12}$ cm$^{-2}$.  If the density
drops by a factor of two (whereas all the other parameters are fixed),
the \NTHP (1--0) excitation temperature drops to 3.6 K.  In these
conditions, using T$_{\rm ex}$ = 4.4 K instead of 3.6 K, in our
analytic column density determination (see Sect.~\ref{GASCOLUMN}),
implies underestimating N(\NTHP) by 50\%.  Therefore, our assumption
of constant T$_{\rm ex}$ can be the main cause of the observed f$_{\rm
D}$ scatter at low extinctions.

Because of the anti-correlation between dust temperature and column
density (see Fig.~\ref{TDVSAV}), we expect that there will also be an
anti-correlation between the depletion factor, $f_D$, and dust
temperature.  Figure \ref{IRAMFDNH} (right panels) shows the depletion
factor for each molecule plotted against the line-of-sight averaged
dust temperature.  As expected, the depletion is highest in the low
temperature regions though the lower signal to noise in \cs\ and
\dcop\ make this somewhat harder to see.  The anti-correlation between
the depletion factor and dust temperature has also been seen by
\citet{Kramer99} in IC5146 in \ceo, though in TMC-1C the temperatures
are somewhat lower.

Our data clearly suggest that there is an increasing depletion of
\NTHP \ with increasing \MOLH \ column (and volume) density.  In
previous work (e.g. Tafalla et al. 2002, 2004; Vastel et al. 2006),
the observed \NTHP \ abundance appears constant across the core,
although the data are also consistent with chemical models in which
the \NTHP \ abundance decreases by factors of a few
\citep{Caselli02c}.  \citet{Bergin02} also deduce small depletion
factors for \NTHP \ when comparing data to models and \citet{Pagani05}
found clear signs of \NTHP \ depletions at densities above
$\sim$10$^5$ \percc . There is also evidence of \NTHP \ depletion
towards the Class 0 protostar IRAM 04191+1522 \citep{Belloche04} in
Taurus. The average \NTHP \ abundance that we find in TMC-1C, relative
to H$_2$, is 1.4$\times10^{-10}$.

What appears to be different from previous work is that the CO
depletion factor towards the dust peak is relatively low (compared to,
e.g., L1544), and, {\it at the same time}, \NTHP (1--0) lines are
bright over an extended region. If TMC--1C were chemically young (such
as L1521E; Tafalla \& Santiago 2004, Hirota et al. 2002), then there
would be negligible CO freeze--out and low abundances of \NTHP , given
that \NTHP\ is a ``late-type'' molecule (i.e. its formation requires
significantly longer times (factors $>$10) than CO and other
C--bearing species).  In TMC--1C we observe moderate CO {\it and}
\NTHP depletions, as well as extended \NTHP \ emission with derived
fractional abundances around 10$^{-10}$. To derive an approximate
value of the average gas number density of the region where \NTHP \
emission is present, we first sum all the observed \NTHP (1--0)
spectra (over the whole mapped area, with size $\simeq$450\arcsec
$\times$170\arcsec , corresponding to a linear geometric mean of about
40,000~AU; see Fig.~\ref{COARROWS}) and then perform an hfs fit in
CLASS to derive the excitation temperature.  We find $T_{\rm ex}$ =
3.6$\pm$0.02~K and $N(\NTHP )$ = 4.84$\pm$0.03 $\times 10^{12}$ \cmsq
, which can be reproduced with the LVG code if $n(\MOLH )$ =
5$\times$10$^3$ \percc \ and $T_{\rm kin}$ = 11~K, as found in
previous sections (see Sect.~\ref{INFALL} and \ref{GASTEMP}).  The
average value of the extinction across the whole TMC--1C core is 23
mag, so that the corresponding \NTHP\ abundance is 2$\times 10^{-10}$,
close to the average value found before.  How long does it take to
form \NTHP\ with fractional abundances of $\sim10^{-10}$ in regions
with volume densities $\sim5\times 10^3$ cm$^{-3}$?  \citet{Roberts04}
derive times $\geq 3\times 10^5$ yr at n(H$_2$) = 10$^4$ cm$^{-3}$, so
that this can be considered {\it a lower limit to the age of the
TMC--1C core}.

In summary, all the above observational evidence suggests that the
majority of the gas observed towards TMC--1C has been at densities
$\sim$10$^4$~cm$^{-3}$ for at least a few times 10$^5$ yr and that
material is accreting toward the region marked by the mm dust emission
peak.  We finally note that the density profile of the region centered
at the dust peak position is {\it steeper} (consistent with a power
law; see Fig.~13 of Schnee \& Goodman 2005) than found in other cores,
so that Bonnor-Ebert spheres may not be the unique structure of dense
cores in their early stages of evolution.

\subsection{Chemical Model}
\label{schem}

Although TMC--1C is more massive than L1544 by a factor of about two,
the physical structures of the two cores are similar: the central
density of TMC--1C is $\sim5\times 10^5$ \percc \ (factor of $\sim$ 2
lower than L1544, according to Tafalla et al. 2002) and the central
temperature is $\sim$7~K, similar to the dust temperature deduced by
Evans et al. (2001) and Zucconi et al. (2001) in the center of L1544,
and close to the gas temperature recently measured by Crapsi et
al. (2007), again toward the L1544 center.  However, the chemical
characteristics of the two cores appear quite different.  In TMC--1C:
1. the observed CO depletion factor is about 4.5 times smaller than in
L1544 (see Section \ref{smol} and \citet{Crapsi05}); 2. the deuterium
fractionation is three times lower \citep{Crapsi05} than in L1544;
3. the \NTHP \ column density at the dust peak is two times lower and
the \CSEO \ column density is 1.7 times larger than in L1544
\citep{Caselli02c}.  All this is consistent with a younger chemical
(and dynamical) age \citep{Shematovich03, Aikawa05}.

\begin{figure}
\epsscale{0.7}
\includegraphics[angle=270,width=6.0in]{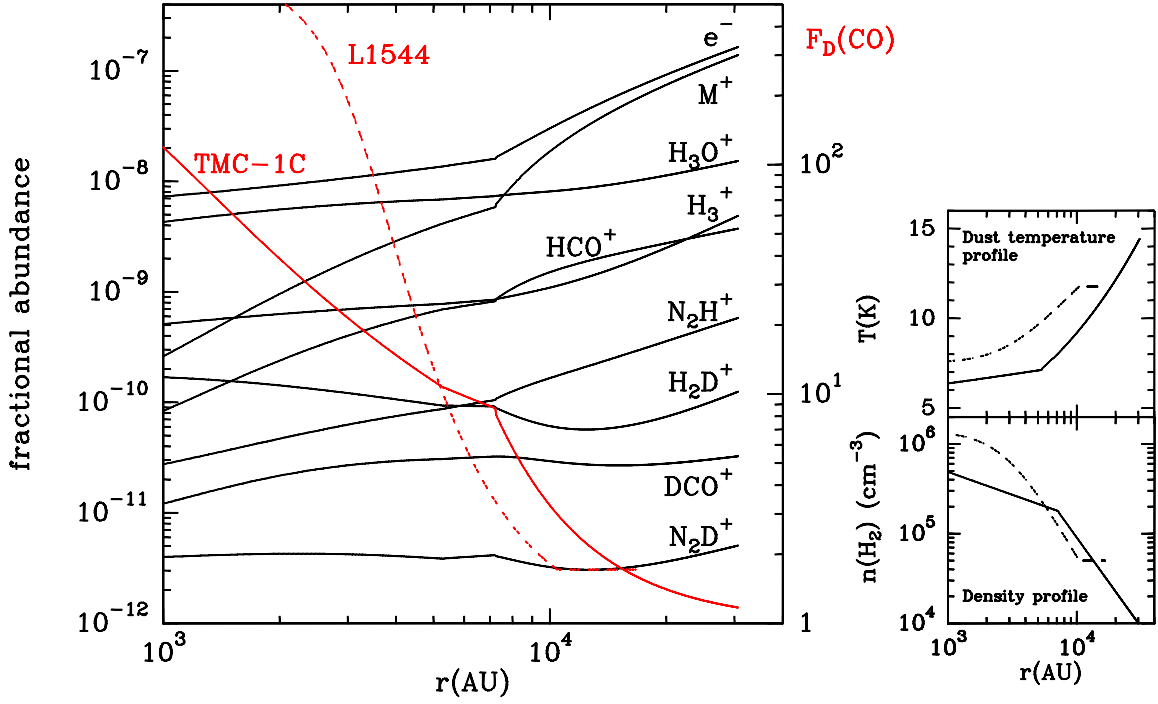}
\caption{({\it Left panel}) Fractional abundances of important
  molecular ions as a function of radius, calculated by the best--fit
  model for TMC--1C (i.e. the one that best reproduces the observed
  column densities, see text).  The red curves are depletion factors
  {\it within} TMC-1C ($F_{\rm D}(CO)$, see the scale in the right
  y-axis) and, for comparison, within L1544 (dashed red curve, from
  Vastel et al. 2006). The discontinuity in the fractional abundance
  profiles is due to the particular density profile used.  ({\it Right
  panel}) Temperature and density profiles in TMC--1C, derived by
  \citet{Schnee06b} and used in the chemical model. Dashed curves
  refer to L1544 (adopted by Vastel et al. 2006). \label{fchem}}
\end{figure}

To understand this chemical differentiation in objects in apparently
similar dynamical phases, we used the simple chemical model originally
described in \citet{Caselli02c} and more recently updated by
\citet{Vastel06}.  The model consists of a spherical cloud with
density and temperature gradients as determined by \citet{Schnee06b}.
The model starts with \MOLH , \MOLN , CO and O in the gas phase, a
gas--to--dust mass ratio of 100, and a Mathis, Rumpl \& Nordsiek
(1977; MRN) grain size distribution. Molecules and atoms are allowed
to freeze--out onto dust grains and desorb via cosmic--ray impulsive
heating \citep{Caselli02c, Hasegawa93}. The adopted binding energies
of CO and \MOLN \ are 1100~K and 982.3~K, respectively.  The CO
binding energy is intermediate between the one measured for CO onto
(i) icy mantles (1180~K; Collings et al. 2003 and Fraser et al. 2004)
and (ii) CO mantles (885~K; \"Oberg et al. 2005).  The adopted value
(1100~K) is the weighted mean of the two measured values, assuming
that water is about four times more abundant than CO in the Taurus
molecular cloud (see Table 2 of Ehrenfreund \& Charnley 2000 and
references therein).  See \citet{Oberg05} for adsorption onto icy
mantles. For the atomic oxygen binding energy we used 750~K, as in
\citet{Vastel06}.  The following parameters have also been assumed
from \citet{Vastel06}: (i) the cosmic ray ionization rate
(1.3$\times$10$^{-17}$ s$^{-1}$); (ii) the minimum size of dust grains
($a_{\rm min}$ = 5$\times10^{-6}$ cm); (iii) the ``canonical''
abundance of CO (9.5$\times$10$^{-5}$, from \citet{Frerking82}; (iv)
the sticking coefficient ($S$=1, as recently found by
\citet{Bisschop06} for CO and \MOLN ); (v) the initial abundance of
\MOLN \ equal to 4$\times 10^{-5}$, i.e. about 50\% the total
abundance of nitrogen observed in the interstellar medium
\citep{Meyer97}; (vi) the initial abundance of ``metals'' (M$^+$, in
Fig.~\ref{fchem}) of 10$^{-6}$ (from \citet{McKee89}); and (vii) the
initial abundance of Oxygen, fixed at a half the canonical abundance
of CO (i.e. 13 times lower than the cosmic abundance; Meyer et al.
1998).

The model is run until the \CSEO \ column density toward the center of
the cloud reaches the observed value ($t$ = 8$\times$10$^3$ yr).
During this time, the abundance of molecular ions is calculated within
the cloud using steady state chemical equations with the instantaneous
abundances of the neutral species. To determine $x(e)$, the reaction
scheme of \citet{Umebayashi90} is used, where the abundance of the
generic molecular ion ``mH$^+$'' (essentially the sum of \HCOP , \NTHP
, \HTHOP \ and their deuterated forms) is calculated (see
\citet{Caselli02c} for more details).  The calculated abundance
profiles of the various species have then been convolved with the
HPBWs of the 30m antenna at the corresponding frequencies and the
derived column densities are in very good agreement with the observed
quantities (within factors of 2 for \NTHP , \NTDP \ and, of course, CO
isotopologues), which is very encouraging, considering the simplicity
of the model.

The best-fit chemical structure of TMC--1C, reached after 10,000 yr,
is shown in the left panel of Fig.~\ref{fchem}.  Note that despite the
similar binding energies of CO and \MOLN , the \HCOP \ and \DCOP \
drops are steeper than those of \NTHP \ and \NTDP , which is due to
the fact that the CO freeze--out (although lower than in L1544)
enhances the \NTHP\ production rate, as pointed out by previous
chemical models \citep{Aikawa01}.  Finally, we note that the CO
depletion factor {\it within} the cloud, $F_{\rm D}$(CO)\footnote{The
symbol $F_{\rm D}$(CO), used here to indicate the CO depletion {\it
within} the cloud, should not be confused with $f_{\rm D}$(CO), the
{\it observed} (or integrated-along-the-line-of-site) CO depletion
factor ($f_{\rm D}$ = $\int F_{\rm D}(l) dl$; see also Crapsi et
al. 2004).}, is significantly lower than in L1544 at radii $\la$ 5,000
AU, which reflects the different density profile (see right bottom
panel in Fig.~\ref{fchem}).

The present data, together with previous work, show that there are
significant chemical variations among apparently similar cloud cores
and that CO is not always heavily depleted when the volume density
becomes larger than a few $\times$10$^4$ \percc \ (as found in L1544,
L1498 and L1517B; Caselli et al. 2002, Tafalla et al. 2002,
2004). Indeed, the case of L1521E, a Taurus starless core, where the
central density is 10$^5$ \percc \ but no CO freeze--out is observed
\citep{Tafalla04b}, suggests that cloud cores in similar dynamical
stages can have different chemical compositions. This point has been
further discussed by \citet{Lee03}, who underline the importance of
the environment in setting the chemical/dynamical stage of a core, so
that a core like L1521E (and L1689B) may have experienced a recent
contraction phase, where the chemistry has not yet had the time to
adjust to the new physical structure.  On the other hand, the Bok
Globule B68, being close to equilibrium, may have achieved the present
structure a long time ago, so that both CO and \NTHP \ had time to
freeze--out (as found by \citet{Bergin02}).

The present detailed study of TMC--1C adds a new piece to the puzzle:
cores which are currently accreting material from the surrounding
cloud appear chemically younger, with lower CO depletion factors.  If
the accreting cloud material, at densities $\la$10$^4$ \percc , is old
enough ($\ga 3\times 10^5$ yr) to have formed observable abundances of
\NTHP\ (as in TMC--1C), then \NTHP \ lines toward the dust peak will
be bright. On the other hand, the chemical structure of cores such as
L1521E (rich in CO, but poor in N--bearing species such as \NTHP \ and
\AMM ) may be understood as young condensations which are accreting
either lower density material (where the chemical times scales for
\NTHP\ formation are significantly longer) or material which spent
only a small fraction of 10$^5$ yr at relatively high densities
($\sim$10$^4$ \percc ). The former hypothesis may be valid in
environments less massive than those associated with TMC--1C (and it
does not require large contraction speeds), whereas the latter
hypothesis needs dynamical time scales shorter than or at most
comparable to chemical time scales (which are about 10$^4$ yr at
densities of $\sim$10$^5$ \percc , as can be found from the
freeze--out time scale of species such as CO).  

In summary, TMC-1C, being more massive than L1544 and other typical
low-mass cores, has a {\it larger reservoir} of undepleted material at
densities close to 10$^4$\percc , where both CO and \NTHP\ are
abundant.  Detailed chemical models suggest that TMC--1C must be at
least $3\times 10^5$~yr old, to reproduce the observed \NTHP\
abundances across the cloud. Our simple chemical code tell us that the
observed CO depletion factors can be reached in only 10,000 yr.
Therefore, the core nucleus is either significantly younger than the
surrounding material or the surrounding (undepleted) material has
accreted toward the core nucleus in the past 10,000 yr.  In either
case, this is evidence that {\it the densest part of TMC--1C has
recently accreted material}.

From the velocity gradients presented in Section \ref{svel} it is hard
to see a clear pattern of flowing material towards the dust peak, but
the ``chaotic'' pattern is reminiscent of a turbulent flow that may be
funnelling towards the densest region, aided by gravity.  In any case,
the extended inward motions deduced from \NTHP\ observations (see
Sect.~\ref{INFALL}) is consistent with material at about
(5--10)$\times$10$^3$ \percc \ {\it currently accreting} toward the
dust peak position at velocities around 0.1 \kms .  It will be
extremely important to compare this velocity field with those
predicted by turbulent simulations of molecular cloud evolution,
especially considering the possibility of competitive accretion
(Bonnell \& Bate 2006).  Observation of CS will also be important to
check our prediction of a larger ``extended infall'' velocity, when
compared to L1544.

\section{Summary} \label{SUMMARY}

A detailed observational study of the starless core TMC--1C, embedded
in the Taurus molecular cloud, has been carried out with the IRAM 30m
antenna.  We have determined that TMC-1C is a relatively young core (t
$\ga 3\times 10^5$ yr), with evidence of material accreting toward the
core nucleus (located at the dust emission peak).  The core material
at densities $\ga$10$^5$ \percc \ is embedded in a cloud condensation
with total mass of about 14 M$_{\odot}$ and average density of
$\simeq$10$^4$ \percc , where CO is mostly in the gas phase and \NTHP
\ had the time to reach the observed abundances of $\simeq$10$^{-10}$.
The overall structure is suggestive of ongoing inflow of material
toward the central condensation.  In addition, we have found that:

\noindent
1. \NTHP (1--0) lines show signs of inward asymmetry over a region of
about 7,000 AU in radius. This is the most extended inward asymmetry
observed in \NTHP \ so far. The data are consistent with simple
two--layer models, where the line-of-sight component of the relative
(infall) velocities range from $\sim$0.15 \kms \ (toward the dust
peak) to $\sim$0.05 \kms \ (at a distance from the dust peak of about
7,000 AU).

\noindent
2. CO isotopologues and \NTHP \ show increasing depletion as $A_V$
increases and $T_d$ decreases.  The amount of CO depletion that we
observe is a factor of $\sim$5 lower than that of L1544, whereas \NTHP
\ column densities are only a factor of two lower. Also, \NTHP \ show
clear signs of moderate depletion toward the dust peak position.

\noindent
3. The gas temperature determined from \CEIO (2--1) is 12~K at the
dust peak, indicating that CO is not tracing the dense ($n(\MOLH) > 5
\times 10^4$ \percc) and cold ($T_{\rm kin} <$ 10~K) regions of dense
cores. The \CEIO (2--1) excitation temperature drops outside the dust
peak, and this is consistent with a roughly constant kinetic
temperature and a dropping volume density ({\it traced by CO
isotopologues}) from $\sim 4 \times 10^4$\percc\ toward the dust peak
to $\simeq 6 \times 10^3$\percc\ at a projected distance from the dust
peak of about 11,000~AU (no high S/N data are available to probe
larger size scales).

\noindent
4.  \nthp (1-0) line widths are constant across the core, which is
consistent with previous NH$_3$ measurements \citep{Barranco98}, but
different from what has been found with \nthp\ and \ntdp\ observations
of L1544 and L1521F \citep{Crapsi05}, where line widths are increasing
toward the core center.  The increase in line width with radius seen
in \ceo\ and \nthp\ in B68 \citep{Lada03} is not seen in TMC--1C, and
unlike B68, the \ceo\ line width is significantly larger than the
\nthp\ line width throughout TMC-1C. This is consistent with 
the fact that TMC--1C, unlike B68, is embedded in a molecular
cloud complex, so that CO lines trace more material along the 
line of sight of TMC--1C. 

\noindent
5. The velocity field that we see in TMC--1C does not show the global
signs of rotation that were seen in NH$_3$ observations over a
somewhat different area at arcminute resolution in \citet{Goodman93}.
Nevertheless, one portion of TMC--1C encompassing the dust peak
position does have a more coherent velocity field, suggestive of solid
body rotation with magnitude $\simeq$4 \kms pc$^{-1}$, in the tracers
\cso (1-0), \cso (2-1), \ceo (2-1) and \NTHP (1-0).

\noindent
6. The observed chemical structure of the TMC--1C core can be
reproduced with a simple chemical model, adopting the CO and N$_2$
binding energies recently measured in the laboratory.  We argue here
that ``chemically young and physically evolved'' cores like L1521E and
L1698B (those with low CO depletion, faint \NTHP\ lines, central
densities above $10^5$ cm$^{-3}$ and centrally concentrated structure)
have lower density envelopes than TMC--1C in which the \NTHP\
abundance did not have the time to reach equilibrium values.  On the
other hand, ``chemically and physically evolved cores'' like L1544,
L694-2 and L183 (those with high CO depletion and bright \NTHP\ lines)
are likely to have lower rates of accretion of material from the
envelope to the nucleus than in TMC--1C (or it has ended), and with a
core nucleus undergoing contraction.  Finally, ``chemically evolved''
but less centrally concentrated cores (e.g. L1498, L1512, B68), can
just be older objects (age $\ga 10^6$ yr), close to equilibrium, as
suggested by Lada et al. (2003).  In the case of TMC--1C, there is
evidence that the core is at least $3\times10^5$ yr old and has
recently accreted less chemically evolved material.

More comprehensive chemical models, taking into account the accretion
of chemically young material, as well as a comparison between the
observed velocity patterns and turbulent models of cloud core
formation are sorely needed to test our conclusions.

\acknowledgments

Our anonymous referee has provided valuable comments and suggestions
which have improved the content and clarity of this paper.  We would
like to thank Phil Myers, Ramesh Narayan, David Wilner and Doug
Johnstone for their suggestions, assistance, and insights.  The James
Clerk Maxwell Telescope is operated by The Joint Astronomy Centre on
behalf of the Particle Physics and Astronomy Research Council of the
United Kingdom, the Netherlands Organisation for Scientific Research,
and the National Research Council of Canada.  IRAM is supported by
INSU/CNRS (France), MPG (Germany), and IGN (Spain).  This material is
based upon work supported under a National Science Foundation Graduate
Research Fellowship.

\appendix

\section{Error estimates on $T_{\rm ex}$}

From the equation of radiative transfer (see eq.~\ref{erad}), once
$T_{\rm mb}$, $\tau$ and the corresponding errors are known, the error
on $T_{\rm ex}$ ($\sigma_{T_{\rm ex}}$) can be determined following
the rules of error propagation:
\begin{eqnarray}
\sigma_{T_{\rm ex}}^2 & = & \left( 
\frac{\partial T_{\rm ex}}{\partial T_{\rm mb}} \sigma_{T_{\rm mb}} \right)^2
+ \left( \frac{\partial T_{\rm ex}}{\partial \tau} \sigma_{\tau} \right)^2 ,
\label{esig}
\end{eqnarray}
where $\sigma_{T_{\rm mb}}$ and $\sigma_{\tau}$ are the errors
associated with $T_{\rm mb}$ and $\tau$, respectively.

The expression of $T_{\rm ex}$ is found by inverting eq.~\ref{erad}:
\begin{eqnarray}
T_{\rm ex} & = & \frac{T_0}{\ln [T_0/A + 1]} \, , \, {\rm where} \\ \nonumber
A & = & \frac{T_{\rm mb}}{1 - e^{-\tau}} + J_{\nu}(T_{\rm bg})  .
\end{eqnarray}
Thus, the partial derivatives in eq.~\ref{esig} are:
\begin{eqnarray}
\frac{\partial T_{\rm ex}}{\partial T_{\rm mb}} & = & 
\frac{a^2 b}{(bc + T_{\rm mb}) (ab+cb+T_{\rm mb}) \ln \left( 
\frac{ab}{bc + T_{\rm mb}} + 1 \right)^2} \\
\frac{\partial T_{\rm ex}}{\partial \tau} & = & 
- a \ln \left[ \frac{a}{c - T_{\rm mb}/b} + 1 \right]^2 
\times \frac{a T_{\rm mb} e^{\tau}}{(e^{\tau} T_{\rm mb} -
b c)(e^{\tau} T_{\rm mb} - a b -b c)} 
\end{eqnarray}
where,
\begin{eqnarray}
a & \equiv & T_0 \\
b & \equiv & 1 - e^{-\tau} \\
c & \equiv & J_{\nu}(T_{\rm bg}) .
\end{eqnarray}

\clearpage
\begin{deluxetable}{lcccccc} 
\tablewidth{0pt}
\tabletypesize{\scriptsize}
\tablecaption{Observing Parameters of IRAM Spectra \label{IRAMOBS}}
\tablehead{
 \colhead{Transition}		& \colhead{$T_{sys}$}		&
 \colhead{Spectral Resolution}	& \colhead{FWHM}		&
 \colhead{$B_{eff}$}		& \colhead{Frequency}           &
 \colhead{VLSR}                \\
 \colhead{}			& \colhead{Kelvin}		&
 \colhead{\kms}			& \colhead{arcseconds}		&
 \colhead{}                     & \colhead{GHz}                 &
 \colhead{\kms}}		
\startdata
 \cso  (1-0) &  353 & 0.052 & 18.4 & 0.66 & 112.3592837\tablenotemark{a} & 5.2 \\
 \cso  (2-1) & 1893 & 0.052 &  9.2 & 0.40 & 224.7143850\tablenotemark{a} & 5.2 \\
 \ceo  (2-1) & 1050 & 0.053 &  9.4 & 0.41 & 219.5603541\tablenotemark{a} & 5.2 \\
 \cs   (2-1) &  202 & 0.061 & 21.4 & 0.72 &  96.4129495\tablenotemark{a} & 5.2 \\
 \dcop (2-1) &  479 & 0.041 & 14.3 & 0.54 & 144.0773190\tablenotemark{a} & 5.2 \\
 \dcop (3-2) &  622 & 0.054 &  9.5 & 0.42 & 216.1126045\tablenotemark{a} & 5.2 \\
 \nthp (1-0) &  185 & 0.031 & 22.1 & 0.73 &  93.1737725\tablenotemark{b} & 5.2 
\enddata
\tablenotetext{a}{Frequency taken from the  Leiden Atomic and 
	          Molecular Database \citep{Schoier05}}
\tablenotetext{b}{Frequency taken from \citet{Dore04}}
\end{deluxetable}

\begin{deluxetable}{lrrrr}
\tablewidth{0pt}
\tabletypesize{\scriptsize}
\tablecaption{Fit Line Parameters at Dust Peak \label{FITTABLE}}
\tablehead{
 \colhead{Transition}                  & \colhead{VLSR}                &
 \colhead{$\int$ Tdv\tablenotemark{a}} & \colhead{FWHM}                &
 \colhead{rms}                 \\
 \colhead{}                            & \colhead{\kms}                &
 \colhead{K \kms}                      & \colhead{\kms}                &
 \colhead{K}}
\startdata 
 \cso  (1-0) & 5.33 $\pm$ 0.03 & 0.37 $\pm$ 0.07 & 0.45 $\pm$ 0.11 & 0.089 \\
 \cso  (2-1) & 5.49 $\pm$ 0.01 & 0.71 $\pm$ 0.05 & 0.43 $\pm$ 0.04 & 0.144 \\
 \ceo  (2-1) & 5.20 $\pm$ 0.01 & 2.07 $\pm$ 0.01 & 0.41 $\pm$ 0.01 & 0.250 \\
 \cs   (2-1) & 5.26 $\pm$ 0.04 & 0.17 $\pm$ 0.03 & 0.50 $\pm$ 0.10 & 0.092 \\
 \dcop (2-1) & 5.14 $\pm$ 0.01 & 0.28 $\pm$ 0.03 & 0.16 $\pm$ 0.02 & 0.204 \\
 \dcop (3-2) & 5.25 $\pm$ 0.02 & 0.34 $\pm$ 0.04 & 0.26 $\pm$ 0.04 & 0.179 \\
 \nthp (1-0) & 5.20 $\pm$ 0.01 & 0.20 $\pm$ 0.03 & 0.19 $\pm$ 0.03 & 0.104 
\enddata
\tablenotetext{a}{Integrated intensity over entire spectrum (all hyperfine
components included in the case of \CSEO \ and \NTHP ).}
\end{deluxetable}

\begin{deluxetable}{ccccc}
\tablewidth{0pt}
\tabletypesize{\scriptsize}
\tablecaption{Fit results for the five spectra in Fig.~\ref{two_layer} 
\label{table_layer}}
\tablehead{
 \colhead{Offset}           & \colhead{$T_{\rm ex}$}  &
 \colhead{$\tau_{\rm TOT}$} & \colhead{$\Delta v$}               &
 \colhead{$V_f-V_b$}     \\
 \colhead{(arcsec)}             & \colhead{K}            &
 \colhead{}                  & \colhead{\kms }                    &
 \colhead{\kms }  }
\startdata
-40,60 B\tablenotemark{a} & 4.5 & 17 & 0.25 & 0.0 \\
-40,60 F\tablenotemark{b} & 3.3 & 10 & 0.15 & -0.05 \\
-20,40 B  & 4.5 & 15 & 0.25 & 0.0 \\ 
-20,40 F  & 3.5 & 15 & 0.20 & 0.05 \\
0,20 B    & 4.4 & 23 & 0.17 & 0.0 \\ 
0,20 F    & 3.3 & 10 & 0.23 & 0.15 \\
20,0      & 4.4$\pm$0.3 & 23$\pm$2 & 0.17$\pm$0.01 & ... \\
40,-20    & 4.9$\pm$0.3 & 8.8$\pm$0.8 & 0.22$\pm$0.01 & ... 
\enddata
\tablenotetext{a}{B$\equiv$background layer}
\tablenotetext{b}{F$\equiv$foreground layer}
\end{deluxetable}

\begin{deluxetable}{lrrrrr} 
\tablewidth{0pt}
\tabletypesize{\scriptsize}
\tablecaption{Molecular Transition Constants and Assumed Abundances\label{CONSTANTS5}}
\tablehead{
 \colhead{Transition}		& \colhead{A\tablenotemark{a}}	&
 \colhead{B\tablenotemark{b,c}}	& \colhead{$g_l$}		&
 \colhead{$g_u$}		& \colhead{$X_{can}\tablenotemark{d}$} \\
 \colhead{}			& \colhead{$s^{-1}$}		&
 \colhead{GHz}			& \colhead{}	    		&
 \colhead{}                     & \colhead{$n/n_{H_2}$}}
\startdata
 \cso  (1-0) & 6.697E-8 & 56.179990 & 1 & 3 & $4.7 \times 10^{-8}$\tablenotemark{e}  \\
 \cso  (2-1) & 6.425E-7 & 56.179990 & 3 & 5 & $4.7 \times 10^{-8}$\tablenotemark{e}  \\
 \ceo  (2-1) & 6.011E-7 & 54.891420 & 3 & 5 & $1.7 \times 10^{-7}$\tablenotemark{f}  \\ 
 \cs   (2-1) & 1.600E-5 & 24.103548 & 3 & 5 & $1.3 \times 10^{-10}$\tablenotemark{g} \\
 \dcop (2-1) & 2.136E-4 & 36.01976  & 3 & 5 & $2.8 \times 10^{-10}$\tablenotemark{h} \\
 \dcop (3-2) & 7.722E-4 & 36.01976  & 5 & 7 & $2.8 \times 10^{-10}$\tablenotemark{h} \\
 \nthp (1-0) & 3.628E-5 & 46.586867 & 1 & 3 & $1.4 \times 10^{-10}$\tablenotemark{i}
\enddata
\tablenotetext{a}{Einstein A coefficients are taken from the 
		  Leiden Atomic and Molecular Database \citep{Schoier05}}
\tablenotetext{b}{\citep{Frerking82}}
\tablenotetext{c}{\citep{Gottlieb03}}
\tablenotetext{d}{Standard molecular abundance, taken from literature except for \nthp}
\tablenotetext{e}{\citep{Crapsi04}}
\tablenotetext{f}{\citep{Goldsmith97}}
\tablenotetext{g}{\citep{Tafalla02}}
\tablenotetext{h}{\citep{Lee03}}
\tablenotetext{i}{Average value derived across TMC--1C.}
\end{deluxetable}

\begin{deluxetable}{lrrr}
\tablewidth{0pt}
\tabletypesize{\scriptsize}
\tablecaption{Gas Column Density and Depletion at Dust Peak
\label{FDTABLE}}
\tablehead{
 \colhead{Transition}             & \colhead{N}
&
 \colhead{$f_D$}\tablenotemark{a} & \colhead{Percent
Error}\tablenotemark{b} \\
 \colhead{}                       & \colhead{cm$^{-2}$} &
 \colhead{}                       & \colhead{}}
\startdata
 \cso  \tablenotemark{c} & 1.0E15 & 2.8 & 5 \\
 \cso  \tablenotemark{d} & 7.8E14 & 3.6 & 7  \\
 \ceo  \tablenotemark{d} & 2.6E15 & 3.8 & 3  \\
 \cs   \tablenotemark{d} & 5.4E11 & 12  & 20 \\
 \dcop \tablenotemark{d} & 1.3E12 & 10  & 14 \\
 \dcop \tablenotemark{e} & 2.8E12 & 5.9 & 13 \\
 \nthp \tablenotemark{c} & 9.4E12 & 0.9 & 6
\enddata
\tablenotetext{a}{Depletion from dust-derived $N_{H_2} = 5.9 \times
10^{22}$, $N_{tot}$ and $X_{can}$}
\tablenotetext{b}{Error from noise in spectrum, not including $\sim$20\%
calibration uncertainty}
\tablenotetext{c}{Derived from (1-0) transition}
\tablenotetext{d}{Derived from (2-1) transition}
\tablenotetext{e}{Derived from (3-2) transition}
\end{deluxetable}

\clearpage

\end{document}